\documentclass[12pt,a4paper]{article}

\usepackage[applemac]{inputenc}
\usepackage[english]{babel}
\usepackage[T1]{fontenc}

\usepackage{amsmath}
\usepackage{amsfonts}
\usepackage{amssymb}
\usepackage{amsthm}
\usepackage{appendix}
\usepackage{graphicx}
\usepackage{graphics}
\DeclareGraphicsExtensions{.jpg,.ps,.pdf,.png,.eps}
\usepackage{color}
\usepackage{booktabs}
\usepackage[pagebackref]{hyperref}
\usepackage{multirow}
\usepackage{url}

\usepackage {fancyhdr}
\usepackage{lastpage}
\usepackage{comment}
\usepackage {stmaryrd}

\usepackage[square, numbers]{natbib}

\usepackage{pdftricks}
\begin{psinputs}
   \usepackage{pstricks}
   \usepackage{multido}
\end{psinputs}

\usepackage[top=2.5cm, bottom=3cm, left=3cm , right=3cm]{geometry}

\usepackage{listings}

\usepackage{float}

\newtheorem{thm}{Theorem}

\newtheorem{prop}[thm]{Proposition}

\newtheorem{rem}{Remark}

\newtheorem{preuve}{Proof}

\def\E{{\mathbb{E}}}
\def\R{{\mathbb{R}}}

\def\P{{\mathbb{P}}}
\def\Q{{\mathbb{Q}}}

\def\1{{\mathbf{1}}}
\def\F{{\mathcal{F}}}

\def\d{{\mathrm{d}}}

\def\la{{\langle}}
\def\ra{{\rangle}}

\def\tPS{{\widetilde{PS}}}

\def\txi{{\tilde{\xi}}}

\def\bit{\begin{itemize}}
\def\eit{\end{itemize}}

\def\tA{{\tilde{A}}}
\def\tB{{\tilde{B}}}

\newcommand{\rmi}{{\rm (i) $\>\>$}}

\newcommand{\rmii}{{\rm (ii) $\hspace{1.5mm}$}}
\newcommand{\rmiii}{{\rm (iii)$\>\>$}}

\def\bc{\begin{center}}
\def\ec{\end{center}}

\def\bcom{}

\def\edoc{\end{document}}

\usepackage{natbib}
 \bibpunct[, ]{(}{)}{,}{a}{}{,}%
\title{Fast calibration of the Libor Market Model with  Stochastic Volatility and Displaced Diffusion}%
\author{{\small Laurent Devineau$^1$, Pierre-Edouard Arrouy$^2$, Paul Bonnefoy$^3$, Alexandre Boumezoued$^4$}\\
{\small Milliman R\&D$^5$}
}

\begin{document}

\maketitle

\footnotetext[1]{Email: \href{laurent.devineau@milliman.com}{laurent.devineau@milliman.com}}
\footnotetext[2]{Email: \href{pierre-edouard.arrouy@milliman.com}{pierre-edouard.arrouy@milliman.com}}
\footnotetext[3]{Email: \href{paul.bonnefoy@milliman.com}{paul.bonnefoy@milliman.com}}
\footnotetext[4]{Email: \href{alexandre.boumezoued@milliman.com}{alexandre.boumezoued@milliman.com}}
\footnotetext[5]{Milliman, 14 Rue Pergol\`ese, 75016 Paris, France.}

\bigskip


\begin{abstract}
This paper demonstrates the efficiency of using Edgeworth and Gram-Charlier expansions in the calibration of the Libor Market Model with  Stochastic Volatility and Displaced Diffusion (DD-SV-LMM). Our approach brings together two research areas; first, the results regarding the SV-LMM since the work of  \cite{WU2006}, especially on the moment generating function, and second the approximation of density distributions based on Edgeworth or Gram-Charlier expansions. By exploring the analytical tractability of moments up to fourth order, we are able to perform an adjustment of the reference Bachelier model with normal volatilities for skewness and kurtosis, and as a by-product to derive a smile formula relating the volatility to the moneyness with interpretable parameters. As a main conclusion, our numerical results show a 98\% reduction in computational time for the DD-SV-LMM calibration process compared to the classical numerical integration method developed by \cite{HESTON1993}.
\end{abstract}


\section{Introduction}

Our work is motivated by the need in the insurance and banking industry to perform repeated calibrations of financial models. So-called market consistent forecasts are notably required for a variety of topics faced by insurance companies, such as the projection of insurance assets and liabilities, the computation of the Solvency Capital requirement through Nested Simulations, see \cite{DEVINEAU2009} and \cite{BAUER2012}, the implementation of intensive recalibration process within a Least Squares Monte Carlo framework, see \cite{DEVINEAU2013}, as well as for the hedging of Variable Annuities and the computation of trading grids. 
Among the financial models required, those dedicated to interest rates have reached a significant complexity within the insurance market practice compared to those dedicated to other financial drivers, such as stocks and inflation.
Our general purpose relates to the improvement of the calibration procedure of the so-called  LIBOR Market Model with Stochastic Volatility, denoted SV-LMM, which is now widely used as it has proven its ability to reproduce volatility smile and fit market prices in a satisfactory way.
Additionally, in a very low interest rate regime, the use of a displacement coefficient allowing to forecast interest rates in the negative region is becoming a market standard, leading us to study the Displaced Diffusion SV-LMM, denoted DD-SV-LMM in what follows.
In this context this is crucial to get fast calibration procedures, especially when the displacement coefficient itself is included in the calibration process, as such studies require to perform intensive recalibration of this coefficient in order to avoid optimization pitfalls.

Starting from the LIBOR Market Model, \cite{JOSHI2003} extended this framework to both stochastic volatility and displaced diffusion, whereas \cite{WU2006} proposed a version of the stochastic volatility component which is now widely used; on this basis they provided several analytical results such as integral-based formulas for caplets and swaptions. Several other versions of the SV-LMM have been developed in the literature, whose differences mainly lie in the way of modelling the stochastic volatility component and the scope of instruments to be addressed; for other versions of the model, we refer to references in \cite{BRIGO2007}.

Due to the need for intensive repeated calibration of the model, there is a huge interest in overcoming the not-so-fast and sometimes unstable existing calibration procedures. In \cite{WU2006}, pricing under the SV-LMM is performed  based on both the classical \cite{HESTON1993} numerical integration method and the famous Fast Fourier Transform (FFT) approach of \cite{CARR1999}, which has become a standard for option valuation for models with known characteristic function, as it is particularly the case for affine diffusion processes. Although the FFT method leads to a slight reduction (29\%) in computational time compared to the Heston approach in the specific \cite{WU2006} pricing example on a strike grid (see Table 4), both methods rely on numerical integration in the complex field, which is known to embed some numerical instabilities, as already highlighted in \cite{KAHL2005} and  \cite{ALBRECHER2006}  on the example of the Heston model. Additionally, the numerical cost shown by both methods makes repeated calibration procedures out of reach in a reasonable operational time.

To address this issue and propose a more efficient calibration method for the DD-SV-LMM, the aim of this paper is to bring together two research areas; first, the results regarding the SV-LMM since the work of  \cite{WU2006}, especially on the moment generating function, and second the use of density distribution approximation based on Edgeworth and Gram-Charlier expansions. Although an analytical expression for the moment generating function does not exist for the SV-LMM in the general setting, for piecewise constant input parameters however (which are natural in the general practice), recursive closed-forms can be given, see \cite{WU2006}, Proposition 4.1. This is our purpose to take advantage of this analytical tractability and implement expansions avoiding as much as possible numerical derivation and integration. This way, we perform the analytical derivation of moments up to fourth order, based on an analytical differentiation of the moment generating function. This allows us to fully exploit the potential of Edgeworth and Gram-Charlier expansions, which can be seen as an adjustment of the Bachelier model for skewness and kurtosis.


In this spirit, several contributions proposed to adjust models as primarily the Black-Scholes one for non-normal skewness and kurtosis, to overcome the well known strike price biases embedded in the standard Black-Scholes formula for away-from-the-money options.
%
\cite{JARROW1982} derived an option pricing formula based on an Edgeworth expansion of the log-normal distribution, whereas later on, \cite{CORRADO1996} used a Gram-Charlier expansion of the normal density of log-returns in the same modelling framework. Both papers provided convincing numerical results.
In our setting, we develop expansions based on the reference normal distribution; this has the advantage of providing an extension of the Bachelier model, which is our natural reference setting allowing to quote derivative instruments, as caps and swaptions, in a negative rates context; currently, short term swaption volatilities can no longer be computed in the alternative log-normal framework proposed by the Black model.
Also, \cite{POTTERS1998} worked in the framework of Edgeworth expansions. They used a normal density adjusted for skewness and kurtosis, derived an analytical approximation of the volatility as a function of the cumulants, then directly fitted to the observed volatility smile (instead of prices), in an analysis dedicated to stock derivatives; this contribution is a key source of inspiration for our present study. More recent references addressed the use and/or analysis of expansions for financial models in different contexts, see e.g. \cite{SCHLOGL2013}, \cite{CHATEAU2014} and \cite{HESTON2016}.



By bringing together these two fields, our approach avoids the complexity and robustness issues of numerical integration, while shortening the calibration process in a significant way. A key step in the analytical tractability is the explicit derivation of moments up to fourth order which are used thereafter in the Gram-Charlier and Edgeworth expansions. Under our expansion regime, we moreover derive smile formulas relating the volatility to the moneyness. In addition to a faster calibration procedure, this therefore provides additional insights on key features on the volatility smile based on interpretable parameters. As a main conclusion, our numerical results show a  98\% reduction in computational time in the DD-SV-LMM calibration process compared to the classical Fast Fourier Transform.

Our paper is structured as follows. In Section \ref{section_model}, we briefly sketch the swap rate dynamics underlying the DD-SV-LMM, and then proceed with the study of the moment generating function. Section \ref{section_edgeworth} establishes the swaption pricing formula based on Gram-Charlier and Edgeworth expansions, and provides the related smile formulas. Finally, Section \ref{section_numerical} details our numerical results assessing the efficiency of the proposed calibration method in comparison to the classical Heston approach. The paper ends with some concluding remarks.

\section{Swap rate distribution under the DD-SV-LMM}
\label{section_model}
In this section, we briefly sketch the swap rate dynamics under the Libor Market Model with Stochastic Volatility and Displaced Diffusion, denoted DD-SV-LMM in what follows. 
We then present the approximate swap rate dynamics under our normal volatilities framework and displaced diffusion setting, based on an adaptation of the freezing technique.
Finally, we detail the set of key results on the moment generating function which will be useful to derive the analytical approximations in the next Section \ref{section_edgeworth}. 

Although these derivations are new in this context, we omit the steps of the reasoning which are analogous to those presented in \cite{WU2006}, and we refer the reader to this paper for more details.

\subsection{The DD-SV-LMM framework}

Let  $P(t,T)$ be the zero-coupon bond maturing at time $T>t$ with par value 1. Let us introduce $F_j(t), \; j=1,...,M$ the value at time $t$ of the simply compounded forward rate for a period $[T_j,T_{j+1}]$ with length $\Delta T_j = T_ {j+1}-T_j$. The forward rates and zero-coupon bond prices are related through
\begin{equation*}
F_j(t)=\frac{1}{\Delta T_j} \left( \frac{P(t,T_j)}{P(t,T_{j+1})}-1\right).
\end{equation*}

In a very low interest rate regime, the use of a displacement coefficient allows for modelling and forecasting interest rates in the negative region. Let us introduce the displacement coefficient $\delta\geq 0$, also called shift, and the $\delta$-displaced forward rate $F_j(t)+\delta$.
The displacement coefficient $\delta$ accounts for possibly negative forward rate $F_j(t)$, while allowing for a log-modelling of $F_j(t)+\delta$. Let us introduce the forward measure $\Q^{j+1}$ associated with the numeraire $P(t,T_{j+1})$; under $\Q^{j+1}$, the displaced forward rate follows the dynamics:
\begin{equation}
\label{equation_forward}
\text{ for } t\leq T_j, \;\;\d F_j(t) = (F_j(t)+\delta) \zeta_j(t) \cdot \d Z_t^{j+1},
\end{equation}
where the inner product '$\cdot$' involves a volatility vector $\zeta_j(t)$ and a multi-dimensional Brownian motion under $\Q^{j+1}$, denoted $Z^{j+1}$. In what follows, we denote by $m(t)=\inf\{ j \geq 1 :  t \leq T_{j} \}$ the first forward rate that has not expired by $t$.
In the model,  the stochastic volatility component is specified as $\zeta_j(t)=\sqrt{V(t)} \gamma_j(t)$, where $\gamma_j(t)$ is a deterministic vector and $V(t)$ lies in the family of Cox-Ingersoll-Ross processes under the spot Libor measure $\Q$ associated with the numeraire $B(t)=\frac{P(t,T_{m(t)})}{\prod_{i=0}^{m(t)-1} P(T_i,T_{i+1})}$ (sometimes assimilated to the risk neutral measure):
\begin{equation}
\label{equation_dynamique_V}
\d V(t)= \kappa \left( \theta - V(t) \right) \d t + \epsilon \sqrt{V(t)} \d W_t,
\end{equation}
whose Feller condition $2\kappa \theta > \epsilon^2$ ensures that the process has a stationary distribution and remains strictly positive.
From  Equation (\ref{equation_forward}), it is possible to derive the stochastic dynamics of displaced forward rates under the reference risk neutral measure as, for $t\leq T_{j}$,
\begin{equation}
\label{equation_forward2}
\d F_j(t) = (F_j(t)+\delta) \sqrt{V(t)} \gamma_j(t) \cdot \left(  \d Z_t - \sigma_{j+1}(t) \sqrt{V(t)} \d t\right),
\end{equation}
with
\begin{equation*}
\sigma_{j+1}(t)= - \sum_{k=m(t)}^j \frac{\Delta T_k (F_k(t)+\delta)}{1+\Delta T_k F_k(t) } \gamma_k(t),
\end{equation*}
where $Z$ is a multi-dimensional Brownian motion under $\Q$, and correlation between $Z$ and $W$ is specified through 
\begin{equation}
\label{equation_correlation}
\rho_j(t) \d t=\E \left[ \left( \frac{\gamma_j(t)}{\left\| \gamma_j(t) \right\|} \cdot \d Z_t \right) \d W_t \right].
\end{equation}

\subsection{Swap rate dynamics}

Although our study can be adapted to the calibration of the model on caplets without restriction, we rather consider in this paper the calibration of the DD-SV-LMM on swaption volatilities, as it allows us to take into account correlations between forward rates. To do so, we  revisit the swaption pricing as proposed in \cite{WU2006}, here adapted to our setting. The swap forward rate at time $t$ for the period from $T_m$ to $T_n$ writes 
\begin{equation*}
R_{m,n}(t)=\frac{P(t,T_m)-P(t,T_n)}{B^S(t)},
\end{equation*}
where $B^S(t)=\sum_{j=m}^{n-1} \Delta T_j P(t,T_{j+1})$ is the annuity of the swap (which strictly depends on $m$ and $n$ although we omit the notation for simplicity). As a numeraire, $B^S(t)$ defines the forward swap measure $\Q^S$; then the price at time zero of the payer swaption contract with strike $K$ is given by the following expectation under $\Q^S$:
\begin{equation}
\label{equation_swaption_pricing}
PS(0,K)=B^S(0) \E^S \left[ \max (R_{m,n}(T_m)-K,0) \right].
\end{equation}
Using weights $\alpha_j(t)= \frac{\Delta T_j P(t,T_{j+1})}{B^S(t)}$, the swap rate can be rewritten as $R_{m,n}(t)=\sum_{j=m}^{n-1} \alpha_j(t) F_j(t)$. To value the swaption, the dynamics under $\Q^S$ can then be derived as follows (see \cite{WU2006}, Eq. 3.3):
\begin{equation*}
\begin{split}
\d R_{m,n}(t) &= \sqrt{V(t)} \sum_{j=m}^{n-1} \frac{\partial R_{m,n}(t)}{\partial F_j } (F_j(t)+\delta) \gamma_j(t) \cdot \d Z_t^S,\\
\d V(t) &= \kappa \left( \theta - \txi^S(t) V(t) \right) \d t + \epsilon \sqrt{V(t)} \d W^S_t,
\end{split}
\end{equation*}
with $\txi^S(t)=1+\frac{\epsilon}{\kappa}\sum_{j=m}^{n-1} \alpha_j(t) \sum_{k=m(t)}^j \frac{\Delta T_k (F_k(t)+\delta)\rho_k(t) \left\| \gamma_k(t) \right\|}{1+\Delta T_k F_k(t)}.$
The differential of the swap rate with respect to $F_j$ is moreover given by
\begin{equation*}
 \frac{\partial R_{m,n}(t)}{\partial F_j }= \alpha_j(t) + \frac{\Delta T_j}{1+\Delta T_j F_j(t)} \sum_{k=m}^{j-1} \alpha_k(t) \left( F_k(t) - R_{m,n}(t)  \right).
\end{equation*}
At this point, one faces the complexity of the dynamics, as in particular the forward rates are involved in the drift of the stochastic volatility process $V$. Analogously to \cite{ANDERSEN2000}
, we will proceed with the freezing technique which relies on the assumption of low variability of frozen coefficients. 

Moreover, as we aim to model the swap volatility in a normal framework, we here adapt the freezing technique  by fixing 
$$
w_j(0)=\frac{\partial R_{m,n}(0)}{\partial F_j } (F_j(0)+\delta),
$$ 
instead of $\frac{\partial R_{m,n}(0)}{\partial F_j } \frac{F_j(0)+\delta}{R_{m,n}(0)}$ as it would be the case in a log-normal framework. This way, we are able to approximate the swap rate dynamics as follows:
\begin{equation*}
\begin{split}
\d R_{m,n}(t) &= \sqrt{V(t)} \sum_{j=m}^{n-1} w_j(0) \gamma_j(t) \cdot \d Z_t^S, \; 0\leq t < T_m,\\
\d V(t) &= \kappa \left( \theta - \txi^S_0(t) V(t) \right) \d t + \epsilon \sqrt{V(t)} \d W^S_t,
\end{split}
\end{equation*}
where $\txi^S_0(t)=1+\frac{\epsilon}{\kappa} \sum_{j=m}^{n-1} \alpha_j(0) \sum_{k=m(t)}^j \frac{\Delta T_k (F_k(0)+\delta)\rho_k(t) \left\| \gamma_k(t) \right\|}{1+\Delta T_k F_k(0)}$.\\

In our setting, we develop expansions based on the reference normal distribution; this has the advantage of providing an extension of the Bachelier model, which is our natural reference setting allowing to quote derivative instruments in a negative rates context; currently, short term swaption volatilities can no longer be computed in the alternative log-normal framework associated to the Black model.

In this slightly adapted framework, it would still be possible to perform swaption pricing under the well known method developed by \cite{HESTON1993} based on numerical integration involving the characteristic function, see e.g. Equation (2.13) in \cite{WU2006}. 
However, such approach requires the computation of an integral in the complex field, which is known to embed some possible numerical instabilities, as already highlighted in \cite{KAHL2005} and  \cite{ALBRECHER2006}  on the example of the Heston model. Additionally, the computational complexity involved in the numerical scheme makes repeated calibration processes out of reach in a reasonable operational time.

To address this issue and propose a more efficient calibration method for the DD-SV-LMM, we aim at providing analytical approximations of the swap rate density distribution by means of Edgeworth and Gram-Charlier expansions, leading to an adjustment of the famous Bachelier formula for skewness and kurtosis. Before detailing our expansion approach, we recall and adapt in the next subsection useful results on the moment generating function.

\begin{rem}
When additionally one is interested into computing prices for an extended grid of strikes, the problem can be reformulated into computing a collection of summations to which the famous Fast Fourier Transform (FFT) method by \cite{CARR1999} can be applied, see e.g. Equation (5.3) in \cite{WU2006}. In our study, we use as a basis for comparison of the calibration efficiency the classical method developed by \cite{HESTON1993}, as indeed we will consider a limited number of strikes for out-of-the-money swaptions. As such, benchmarking with the FFT method is out of scope of the present study, and similar comparison results must hold as the orders of magnitude of the computation speed of the FFT and the Heston methods are close, see Table 4 in \cite{WU2006}, and given that our 98\% reduction compared to the Heston approach is significant, see Section \ref{section_numerical} for more details. Finally, it is worth mentioning that our pricing method and smile formulas based on Gram-Charlier and Edgeworth expansions provide analytical approximations which explicitly depend on the moneyness, therefore avoiding the need for any numerical integration, and as a consequence any use of the FFT method.
\end{rem}


\subsection{The moment generating function}
\label{subsection_moment_generating_function}

We present here the analytical results regarding the moment generating function in the normal volatilities framework, in which the underlying variable to characterize is the swap forward rate itself, and in our drifted diffusion setting. Let us denote by $\psi$ the moment generating function of the state variable $R_{m,n} (T_m )$, defined by
\begin{equation*}
	\psi\left( R_{m,n} (t),V(t),t;z \right) =\E^S \left[ e^{zR_{m,n} (T_m )} | \F_t \right], ~~~z\in\R.
\end{equation*}
%
Using the fact that the conditional expectation above is a martingale, then applying Itô’s formula and finally identifying the drift term leads to the so-called  Kolmogorov backward equation
\begin{equation}\label{equation_kolmogorov}
	\frac{\partial \psi}{\partial t}
	+\left(\kappa \theta-\kappa\xi V \right) \frac{\partial \psi}{\partial V}
	+\frac{1}{2}\epsilon^2V \frac{\partial^2 \psi}{{\partial V}^2}
	+\epsilon \rho \lambda V \frac{\partial^2 \psi}{\partial V \partial x}
	+\frac{1}{2}\lambda^2V \frac{\partial^2 \psi}{{\partial x}^2}
	=0,
\end{equation}
with notations 
\begin{equation*}
\xi \equiv \tilde{\xi}^S_0(t), \; \lambda \equiv \left\| \sum_{j=m}^{n-1} w_j(0) \gamma_j(t) \right\| \; \text{ and } \; \rho = \frac{1}{\lambda} \sum_{j=m}^{n-1} w_j(0) \left\| \gamma_j(t) \right\| \rho_j(t),
\end{equation*}
and terminal condition
$
	\psi\left( x,V,T_m;z \right)=e^{zx}
$.
Let us remark that this equation differs from the one exhibited in \cite{WU2006} as in the normal volatilities framework, we directly focus on the underlying process $R_{m,n}$ instead of $\ln\left(R_{m,n}+\delta\right)$. For this reason the term $-\frac{1}{2}\lambda^2V \frac{\partial \psi}{\partial x}$ which would appear by applying Itô’s lemma  to the process $\ln\left(R_{m,n}+\delta\right)$ vanishes in Equation (\ref{equation_kolmogorov}). 
Adapting \cite{HESTON1993} to our context, one gets a separable form solution, with notation $\tau=T_m-t$, 
\begin{equation}
\label{equation_moment_generating_function}
		\psi\left( x,V,t;z \right)=e^{A(\tau,z)+B(\tau,z)V+zx},
\end{equation}
where
\begin{equation}\label{equation_partial}
	\left\{
		\begin{split}
		 	\frac{\partial A}{\partial \tau}&=\kappa \theta B,\\
			\frac{\partial B}{\partial \tau}&=
				\frac{1}{2}\epsilon^2 B^2
				+\left(\epsilon\rho\lambda z-\kappa \xi \right)B
				+\frac{1}{2}\lambda^2 z^2,
		\end{split}
	\right.
\end{equation}
with boundary conditions $A(0,z)=0$, $B(0,z)=0$. Note that the term $\frac{1}{2}\lambda^2 z^2$ replaces the quantity $\frac{1}{2}\lambda^2\left( z^2-z\right)$ which would appear in a log-normal volatilities framework. 
From \cite{HESTON1993}, it is possible to get an analytical closed-form expression of $A$ and $B$ under the assumption of piece-wise constant functions $\lambda$ and $\rho$ on the grid $(\tau_j,\tau_{j+1}]$, with notation $\tau_j=T_m-T_{m-j}$, which is relevant in practice.
The following recursive backward algorithm allows to compute $A$ and $B$ solution to (\ref{equation_partial}): for each $j=0,...,m-1$, with convention $T_0=0$,
\begin{equation*}
	\left\{
		\begin{split}
		 	A(\tau,z)&=A\left(\tau_j,z\right)+\tA_j(\tau,z)~~~ \forall  \tau \in (\tau_j,\tau_{j+1}], \\
			B(\tau,z)&=B\left(\tau_j,z\right)+\tB_j(\tau,z)~~~\forall  \tau \in (\tau_j,\tau_{j+1}], 
		\end{split}
	\right.
\end{equation*}
where $\tA_j$ and $\tB_j$ are detailed in Appendix \ref{appendix_solving}. 

\section{Swaption pricing and volatility smile derived from Gram-Charlier and Edgeworth expansions}
\label{section_edgeworth}

We present in this section analytical approximations for swaption prices in the DD-SV-LMM framework, allowing to extend the standard Bachelier formula to account for option smiles. The closed-forms rely on Gram-Charlier and Edgeworth expansions at fourth order, which adjust a reference Gaussian distribution by considering skewness and kurtosis. In a first step, we recall some background on Gram-Charlier and Edgeworth expansions, and discuss their main common features and differences. We then derive analytical approximations for swaption prices based on these expansions, and the closed-form derivation of moments of the swap rate up to fourth order. Finally,  we develop smile formulas relating implied volatilities to the moneyness level.

\subsection{Gram-Charlier and Edgeworth expansions}

A Gram-Chalier series expansion (type A) of some density $f$ is defined as
\begin{equation*}
f(z)=\varphi(z) \sum_{n=0}^\infty c_n H_n(z),
\end{equation*}
where $\varphi$ is the standard normal density, the $(c_n)$ are constants related to $f$, and the $(H_n)$ are the Hermite polynomials such that $H_0(z)=1$ and for $n \geq 1$, 
\begin{equation}
\label{equation_hermite}
H_n(z) \varphi(z) = \varphi^{(n)}(z).
\end{equation}
Note that for $i\neq j$, the Hermite polynomials $H_i$ and $H_j$ are orthogonal for the inner product in $L^2(\R)$ defined as $\la F,G\ra=\int_\R F(z) G(z) \varphi(z) \d z$, allowing to identify the coefficients $(c_n)$ which are used in what follows; the proof is left to the reader.

We consider in our study an expansion up to fourth order so as to adjust the reference density for the skewness and kurtosis of the distribution to be estimated, analogously to e.g. \cite{CORRADO1996} and \cite{NECULA2016} where Gram-Charlier series are used to adjust the Black-Scholes formula for equity option prices.  Starting from a random variable $X$ of interest, with standard deviation $\nu$, we consider the density $f$ of the standardized random variable 
\begin{equation}
\label{equation_standardized}
Z=\frac{X-\E[X]}{\nu}.
\end{equation}
Denoting the third and fourth order moments of $Z$ by $\mu_3=\E[Z^3]$ and $\mu_4=\E[Z^4]$ respectively, the fourth order Gram-Charlier approximation, which we denote $g_1$, can be written as
\begin{equation}
\label{equation_gram_charlier}
g_1(z)=\varphi(z) \left\{ 1 - \frac{\mu_3}{6}H_3(z) + \frac{\mu_4-3}{24} H_4(z) \right\}.
\end{equation}

On the other hand, several papers rather focused on Edgeworth-type expansions, see e.g. \cite{BALIEIRO2004}. Originally, and in most applications, Edgeworth expansions are used to provide an approximation of a standardized sum $S_n=\frac{1}{\sqrt{n}}\sum_{j=1}^n X_j$, with $(X_j)$ a sequence of i.i.d. standardized random variables with third and fourth order moments denoted $\gamma_3=\E[X_1^3]$ and $\gamma_4=\E[X_1^4]$ respectively. According to the central limit theorem, the quantity $\P(S_n \leq x)$ converges towards the cumulative distribution function $\Phi(x)=\int_{-\infty}^x \varphi(y) \d y$ of the standard normal distribution. The aim of the Edgeworth expansion is to characterize the distribution of $S_n$ for large $n$. The second order Edgeworth expansion is often considered, which leads to the following approximation using Hermite polynomials introduced in Equation (\ref{equation_hermite}):
\begin{equation*}
\P(S_n \leq x) \approx \Phi(x) - \frac{\gamma_3}{6 \sqrt{n}} \varphi(x) H_2(x) + \frac{\gamma_4-3}{24 n} \varphi(x) H_3(x) + \frac{\gamma_3^2}{72 n} \varphi(x) H_5(x).
\end{equation*}
Note that the term $n$ doesn’t appear in many sudies which aim to derive pricing formulas. This issue is discussed in \cite{BALIEIRO2004}, where the authors indicate that the term $n$ is incorporated to skewness and kurtosis coefficients, and leave these considerations to the reader; we propose to further detail these aspects in Appendix \ref{appendix_edgeworth}.

Let us now provide the approximation based on the single standardized random variable given in Equation (\ref{equation_standardized}) as
\begin{equation*}
\P(Z \leq z) \approx \Phi(z) - \frac{\mu_3}{6} \varphi(z) H_2(z) + \frac{\mu_4-3}{24} \varphi(z) H_3(z) + \frac{\mu_3^2}{72} \varphi(z) H_5(z).
\end{equation*}
Finally, after differentiation, one recovers an Edgeworth approximated density as
\begin{equation}
\label{equation_edgeworth}
g_2(z)=g_1(z)+ \varphi(z) \frac{\mu_3^2}{72} H_6(z),
\end{equation}
where the density $g_1$ is the Gram-Charlier density introduced in (\ref{equation_gram_charlier}). 

\subsection{Swaption pricing}

The standardized random variable of interest is now 
\begin{equation}
\label{equation_Z}
Z=\frac{R_{m,n}(T_m)-R_{m,n}(0)}{\nu},
\end{equation}
with $\nu$ the standard deviation of the swap rate $R_{m,n}(T_m)$. The price of the related swaption given in (\ref{equation_swaption_pricing}) now writes
\begin{equation}
\label{equation_pricing1}
PS(0,K)=B^S(0) \E^S\left[ \max(R_{m,n}(0) + \nu Z -K ,0) \right].
\end{equation}
Let us denote indifferently $g$ the density approximation based on a Gram-Charlier or an Edgeworth expansion, as considered in (\ref{equation_gram_charlier}) and (\ref{equation_edgeworth}) respectively, and still use the notations $\mu_3$ and $\mu_4$ for the third and fourth order moments of $Z$. Let us introduce the standardized moneyness $z_k=\frac{K-R_{m,n}(0)}{\nu}$; then the swaption price given in (\ref{equation_pricing1}) can be approximated by
\begin{equation}
\label{equation_pricing2}
\tPS(0,K)=B^S(0) \int_{z_K}^\infty \left(R_{m,n}(0) + \nu z -K \right) g(z) \d z = \nu B^S(0) \int_{z_K}^\infty \left(z -z_K \right) g(z) \d z.
\end{equation}
Before stating our main result on swaption pricing below, let us recall that the famous Bachelier price $\tPS_0(0,K)$ can be obtained by considering a standard normal distribution for $Z$, leading to
\begin{equation}
\label{equation_bachelier}
\tPS_0(0,K) = \nu B^S(0) \left\{ \varphi(z_K) -  z_K \Phi(-z_K)  \right\},
\end{equation}
where we recall that $\varphi$ and $\Phi$ respectively denote the standard normal density and cumulative distribution function.

\begin{prop}
\label{prop_swaption}
The Gram-Charlier swaption price is given by
\begin{equation}
\label{equation_prop_price_gram}
\tPS_1(0,K) =  \tPS_0(0,K) + \nu B^S(0) \varphi(z_K) \left\{  \frac{\mu_3}{6} z_K + \frac{\mu_4-3}{24} (z_k^2-1) \right\} ,
\end{equation}
and the Edgeworth swaption price writes
\begin{equation}
\label{equation_prop_price_edge}
\tPS_2(0,K) = \tPS_1(0,K) + \nu B^S(0) \varphi(z_K) \frac{\mu_3^2}{72}(z_K^4-6 z_K^2+3).
\end{equation}
According to the Newton binomial formula, the third and fourth order moments of the standardized variable $Z$ defined in Equation (\ref{equation_Z}) are given as follows, for $k\in \{ 3,4 \}$,
\begin{equation*}
\mu_k=\E^S\left[Z^k \right]=\frac{1}{\nu^k} \sum \limits_{j=0}^k \binom{k}{j} \psi^{(j)}(0)\left(-R_{m,n}(0) \right)^{k-j},
\end{equation*}
where $\psi^{(0)}(z)\equiv \psi(R_{m,n}(0),V,0;z)$, with $V\equiv V(0)$, is the moment generating function given in analytical form in Equation (\ref{equation_moment_generating_function}), whose derivatives are given by
\begin{equation}
\label{equation_derivees_Phi}
\left\{
\begin{split}
	\psi^{(1)}&=\left( A^{(1)}_m+B^{(1)}_m V +R_{m,n}(0)\right)\psi^{(0)},\\
	\psi^{(2)} &=
		\left( A_m^{(2)}+B_m^{(2)}V\right) \psi^{(0)}
		+\frac{{( \psi^{(1)})}^2}{\psi^{(0)}},\\
\psi^{(3)}&=
 	\left(  A_m^{(3)}+B_m^{(3)}V\right)\psi^{(0)}
 	+\left(  A_m^{(2)}+B_m^{(2)}V\right)\psi^{(1)}
 	+ \frac{2\psi^{(2)} \psi^{(1)}}{\psi^{(0)}}
 	-\frac{\left(\psi^{(1)}\right)^3}{\left(\psi^{(0)}\right)^2},\\
\psi^{(4)}&=
		 	\left(  A_m^{(4)}+B_m^{(4)}V\right)\psi^{(0)}
		 	+2\left(  A_m^{(3)}+B_m^{(3)}V\right)\psi^{(1)}
		 	+ \left(  A_m^{(2)}+B_m^{(2)}V\right)\psi^{(2)}\\
		 	&+  \frac{2\psi^{(3)} \psi^{(1)}}{\psi^{(0)}}
	 		+ \frac{2 \left(\psi^{(2)}\right)^2}{\psi^{(0)}}
	 		- \frac{5\psi^{(2)}\left(\psi^{(1)}\right)^2 }{\left(\psi^{(0)}\right)^2}
	 		+\frac{2\left(\psi^{(1)}\right)^4}{\left(\psi^{(0)}\right)^3}, 
\end{split}
\right.
\end{equation}
where the computation of the maps $A_m$ and $B_m$ and their derivatives is detailed in Appendix \ref{appendix_moments}.

\end{prop}

Equations (\ref{equation_prop_price_gram}) and (\ref{equation_prop_price_edge}) present the additional terms allowing to adjust the swaptions pricing Bachelier formula by taking into account the skewness and kurtosis of the swap forward rate distribution. Note that the last term of the right-hand side in (\ref{equation_prop_price_edge}) stems from the additional quantity which appears in the Edgeworth expansion compared to the Gram-Charlier’s one, see Equation (\ref{equation_edgeworth}). 
For both formulas, one can check that any distribution with same skewness and kurtosis than the normal distribution are such that $\mu_3=\mu_4-3=0$, which makes the additional terms vanish. 

\begin{rem}
\label{rem_swaptions}
 In the case of at-the-money (ATM) swaptions for which $K=R_{m,n}(0)$, the standardized moneyness $z_K$ is null, so that
\begin{equation*}
\tPS_1(0,R_{m,n}(0)) =  \frac{1}{\sqrt{2 \pi}}  \nu B^S(0) \left\{  1- \frac{\mu_4-3}{24} \right\},
\end{equation*}
and
\begin{equation*}
\tPS_2(0,R_{m,n}(0)) =\frac{1}{\sqrt{2 \pi}}  \nu B^S(0) \left\{  1- \frac{\mu_4-3-\mu_3^2}{24} \right\}.
\end{equation*}
This first shows that even for ATM swaptions, adjusted swaption prices do not match the Bachelier valuation. Furthermore, one can notice that in this case the Gram-Charlier price does not depend on the skewness of the swap rate, whereas the Edgeworth price does through the quantity $\mu_3^2$.
\end{rem}

We now state the proof of Proposition \ref{prop_swaption}.
\begin{preuve}
Let us first note that according to the property (\ref{equation_hermite}) on Hermite polynomials, one can get
\begin{equation*}
\left\{
\begin{split}
H_1(z)&= -z,\\
H_2(z)&= z^2-1, \\
H_3(z)&=-z^3 +3z,\\
H_4(z)&=z^4-6z^2+3.
\end{split}
 \right.
\end{equation*}

Now, let us express the Gram-Charlier density (\ref{equation_gram_charlier}) into the approximated price (\ref{equation_pricing2}), leading to
\begin{equation*}
\begin{split}
\tPS_1(0,K) &=  \nu B^S(0) \int_{z_K}^\infty \left(z -z_K \right) \varphi(z) \d z - \nu B^S(0) \frac{\mu_3}{6} \int_{z_K}^\infty \left(z -z_K \right)  \varphi(z) H_3(z) \d z\\
& + \nu B^S(0) \frac{\mu_4-3}{24}\int_{z_K}^\infty \left(z -z_K \right) \varphi(z) H_4(z)  \d z,
\end{split}
\end{equation*}
where the first component reduces to the Bachelier formula (\ref{equation_bachelier}). 
As for the others, it remains to compute quantities of the following form, for $j=3,4$,
\begin{equation*}
\begin{split}
&\int_{z_K}^\infty \left(z -z_K \right)  \varphi(z) H_j(z) \d z=\int_{z_K}^\infty z  \varphi(z) H_j(z) \d z- z_K \int_{z_K}^\infty   \varphi(z) H_j(z) \d z,
\end{split}
\end{equation*}
By integration by parts and property on Hermite polynomials, see Equation (\ref{equation_hermite}), the first term writes
\begin{equation*}
\int_{z_K}^\infty z  \varphi(z) H_j(z) \d z=- z_K H_{j-1}(z_K) \varphi(z_K) -\int_{z_K}^\infty H_{j-1}(z) \varphi(z) \d z.
\end{equation*}
As for the second term, one gets by the property (\ref{equation_hermite}),
\begin{equation*}
z_K \int_{z_K}^\infty   \varphi(z) H_j(z) \d z = z_K \int_{z_K}^\infty   \varphi^{(j)}(z) \d z = -z_K \varphi^{(j-1)}(z_K) = - z_K H_{j-1}(z_K) \varphi(z_K).
\end{equation*}
This finally leads to
\begin{equation*}
\int_{z_K}^\infty \left(z -z_K \right)  \varphi(z) H_j(z) \d z=-\int_{z_K}^\infty H_{j-1}(z) \varphi(z) \d z = H_{j-2}(z_K) \varphi(z_K),
\end{equation*}
which proves the Gram-Charlier price formula (\ref{equation_prop_price_gram}). The Edgeworth price formula (\ref{equation_prop_price_edge}) can be obtained in a similar way. 
The derivatives in (\ref{equation_derivees_Phi})  of the moment generating function can be derived by standard differentiation of (\ref{equation_moment_generating_function}); this is detailed in Appendix \ref{appendix_moments}.
\end{preuve}

\subsection{Smile formula}

In some calibration frameworks, the underlying target function to minimize in order to estimate the parameters of the DD-SV-LMM is based on volatilities instead of prices. In such a case, it may be useful to consider a smile function rather than inverting theoretical prices with a Bachelier formula. We define by smile function a closed-form expression resulting from the conversion of the Gram-Charlier or Edgeworth prices into an implied Bachelier volatility. The approach we detail hereafter to build such a smile function for swaptions instruments is an adaptation of the method proposed by \cite{BOUCHAUD2003} and \cite{BOUCHAUD2012} for stock implied volatilities.

Let us denote by $s(\nu, z_K)$ the additive correction applied to the volatility $\nu$ in order to recover an implied Bachelier volatility, denoted $\nu(z_K)=\nu+s(\nu,z_K)$. Formally, based on the adjusted volatility, the Bachelier price in (\ref{equation_bachelier}) now writes
$
B^S(0) h\left( \nu + s(\nu, z_K) \right),
$ 
where the function $h$ is given by 
\begin{equation*}
h(x)=\int_{\frac{K-R_{m,n}(0)}{x}}^\infty \big( x z - (K-R_{m,n}(0)) \big) \varphi(z) \d z.
\end{equation*}
The derivative of $h$ at point $\nu$ can be computed as $h'(\nu)=\varphi(z_K)$, which leads to the first order approximation:
\begin{equation}
\label{equation_bachelier_ajuste}
h\left( \nu + s(\nu, z_K) \right) \approx h ( \nu) + s(\nu, z_K) \varphi(z_K).
\end{equation}
On the other hand, one can write the Gram-Charlier and Edgeworth prices in Equations (\ref{equation_prop_price_gram}) and (\ref{equation_prop_price_edge}), which leads to the following result.

\begin{prop}
\label{prop_smile}
The Gram-Charlier smile formula is given by
\begin{equation}
\label{equation_prop_smile_gram}
\nu_1(z_K) =  \nu \left\{  1+ \frac{\mu_3}{6} z_K + \frac{\mu_4-3}{24} (z_k^2-1) \right\},
\end{equation}
and the Edgeworth smile formula writes
\begin{equation}
\label{equation_prop_smile_edge}
\nu_2(z_K)  = \nu_1(z_K)+ \nu \frac{\mu_3^2}{72}(z_K^4-6 z_K^2+3),
\end{equation}
where the third and fourth order moments $\mu_3$ and $\mu_4$ of the swap rate are mentioned in Proposition \ref{prop_swaption} and detailed in Appendix \ref{appendix_moments}.
\end{prop}

\begin{rem}
\label{rem_smile}
 In the case of ATM swaptions, $K=R_{m,n}(0)$, then the standardized moneyness $z_K$ is null, so that the implied volatilities become
\begin{equation*}
\nu_1(0) =  \nu \left\{  1- \frac{\mu_4-3}{24}  \right\} \text{  and  } \nu_2(0) =  \nu \left\{  1 - \frac{\mu_4-3 -\mu_3^2}{24}  \right\}.
\end{equation*}
This shows that the volatility of ATM swaptions $\nu_1(0)$ or $\nu_2(0)$ do not match the forward rate volatility $\nu$. Note in addition that the skewness is only involved in the Edgeworth expansion. These comments are in line with the previous Remark \ref{rem_swaptions} about ATM prices.
\end{rem}

\section{Numerical results}
\label{section_numerical}

This section details the numerical results obtained from the implementation of the expansion methods of Propositions \ref{prop_swaption} and \ref{prop_smile}, and compared to the Heston method illustrated in \cite{WU2006}. We first provide an overview of the calibration setting, including the parametrization of the volatility vector, as well as the market data used. We then compare the Edgeworth and the Gram-Charlier method, and finally compare our approach to the classical Heston method based on numerical integration.

\subsection{Calibration setting}
\label{subsection_calibration_setting}

In our calibration framework, we consider a piecewise constant parametrization of the volatility vector, whose value on the interval $[T_i,T_{i+1})$ is specified as
$
\gamma_j(T_i)=\beta_{j-i+1} g(T_j-T_i)
$
where 
$
g(u)=(bu+a) e^{-cu}+d
$
with non-negative constants $a, b, c \text{ and } d$, and where the $\beta_k$ are 2-dimensional vectors with unitary Euclidian norm.
As for the correlation structure between forward rates and volatilities, we consider a constant parameter $ \rho_j(t) = \rho$, see Equation (\ref{equation_correlation}).
Note that the displaced coefficient $\delta$ is included in the calibration process, so that the set of parameters to be estimated is 
$
\{ a, b, c, d, \kappa, \theta, \epsilon, \rho, \delta \}
$
where we recall that the parameters $\kappa, \theta \text{ and } \epsilon$ are involved in the volatility dynamics, see Equation (\ref{equation_dynamique_V}).

For the purpose of illustration, the market data used for the calibration of the DD-SV-LMM are made of an average interest rate structure and swaption volatilities throughout the year 2016, for both ATM and away-from-the-money swaptions. The ATM swaptions maturities and tenors considered range into $\{1,...,10,15,20,25,30\}$. For away-from-the-money swaption volatilities, we consider the same range for maturities and focus on a 10-years reference tenor; the strikes (in bps) range into +/- $\{25, 50, 100, 150, 200\}$. In the end, this amounts to consider a set of 350 volatilities to replicate.

Finally, note that the target function to be minimized for parameter inference is computed as the sum of squared differences between market and theoretical volatilities.

\subsection{Comparison between the Gram-Charlier and the Edgeworth calibration}
\begin{figure}[h]
\centering
\includegraphics[scale=0.3]{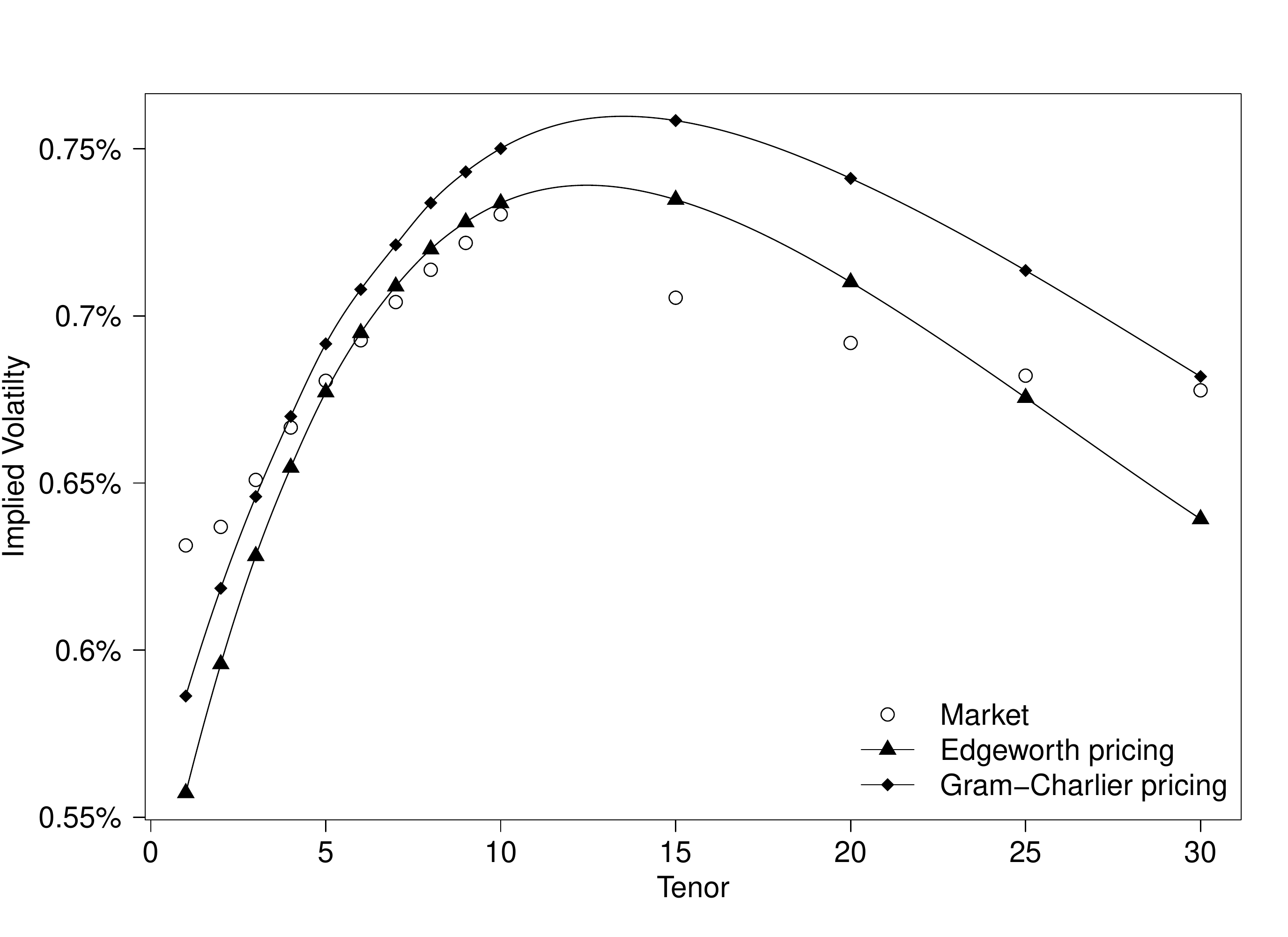}
\caption{ATM Monte Carlo swaption volatilities for 5-years maturity}
\label{figures/graph_EW_GC_5Y_2.pdf}
\end{figure}

\begin{figure}[h]
\centering
\includegraphics[scale=0.3]{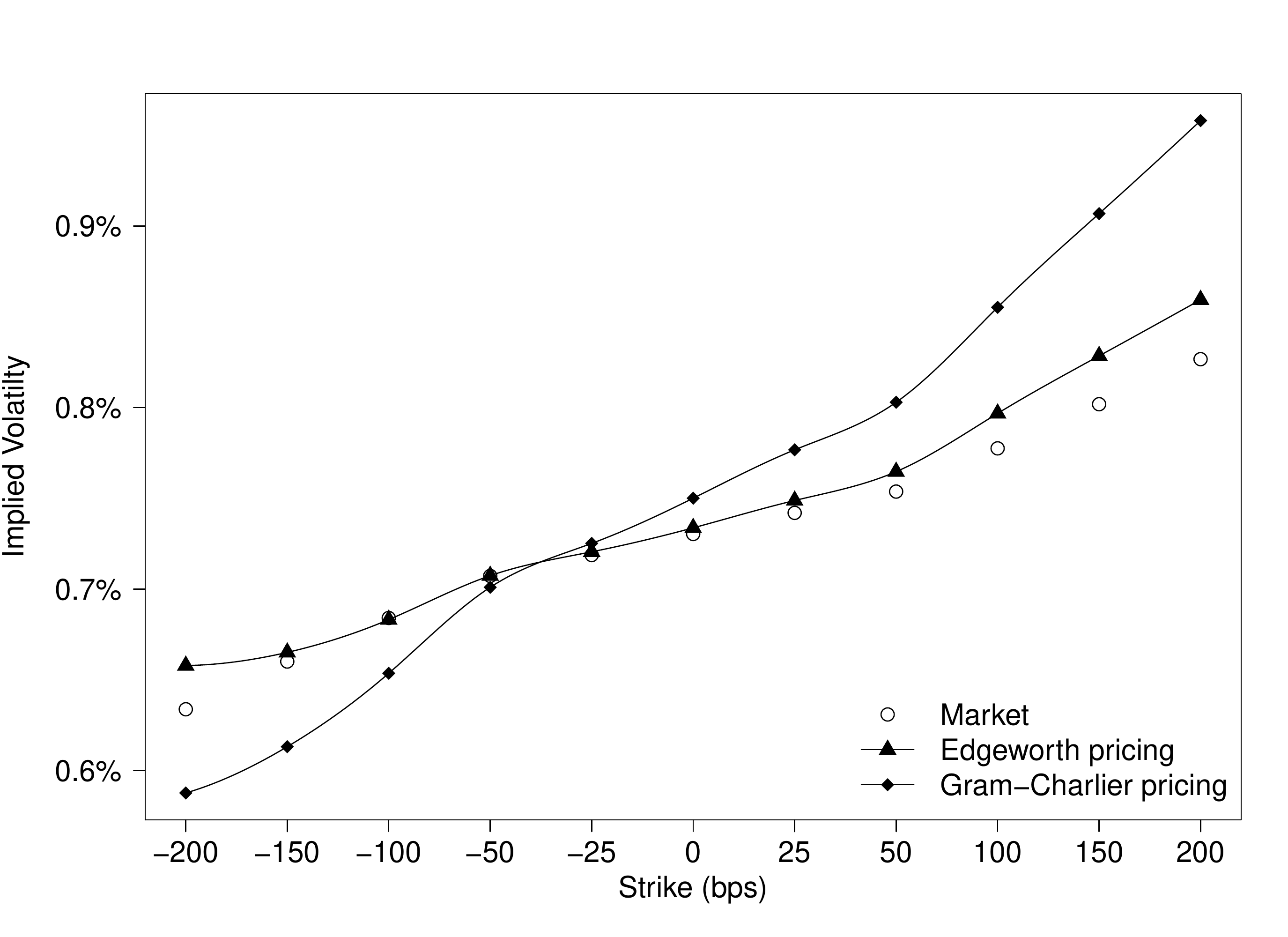}
\caption{Monte Carlo swaption volatility skews for 5-years maturity}
\label{figures/graph_EW_GC_5Y_4.pdf}
\end{figure}

In this part, we discuss the main differences between the Gram-Charlier and the Edgeworth expansions calibration results. As shown in Propositions \ref{prop_swaption} and \ref{prop_smile}, the Edgeworth expansion leads to an additional term compared to Gram-Charlier in the analytical approximation, this term being a function of the skewness of the swap rate distribution. Moreover, for ATM swaptions, the Edgeworth expansion still  accounts for the skewness whereas it vanishes in the Gram-Charlier formulas, see Remarks \ref{rem_swaptions} and \ref{rem_smile}. 
In practice this aspect is illustrated in Figure \ref{figures/graph_EW_GC_5Y_2.pdf} (resp. Figure \ref{figures/graph_EW_GC_5Y_4.pdf}) by the comparison of ATM (resp. away-from-the-money) empirical volatilities for the 5-years maturity.
Empirical volatilities are obtained through the following process: in a first step, we perform a calibration of the DD-SV-LMM; then, forward rates are diffused with the calibrated parameters, and we deduce empirical prices from the Monte Carlo simulations; finally, by inverting the Bachelier formula, we extract empirical volatilities.
The Edgeworth approach shows a better empirical fitting accuracy of market data compared to the Gram-Charlier method. On this basis, we set the Edgeworth approach as the reference one in the next part for the comparison with the classical Heston method.

\subsection{Comparison between the Edgeworth and the Heston methods}

The comparison of our approach with the Heston method illustrated by \cite{WU2006} is analyzed in the light of three criteria:\\
\rmi In a first step, we perform a market consistency analysis assessing the fitting quality of market swaption volatilities, for both theoretical volatilities (implied by the pricing formulas) and empirical volatilities (obtained by Monte Carlo simulation). 
For simulation, a log-Euler scheme is taken with 5000 simulation paths, which reflects an operational standard in the insurance practice, and remains reasonable to provide satisfying convergence of Monte Carlo scenarios. Furthermore, we discuss the skew profile of theoretical volatilities (implied by the pricing formula).\\
\rmii In a second step, we assess the accuracy of the Edgeworth expansion by computing theoretical volatilities based on a common set of parameters.\\
\rmiii In a last step, we present the gain in computational time required for calibration when using Edgeworth expansion, compared to the Heston method detailed in \cite{WU2006}; as a main conclusion, our numerical results show a 98\% reduction in computational time in the DD-SV-LMM calibration process compared to the classical Heston method.

For all three criteria, we present in this section the results focusing on the 5-years maturity; the results for 10-years and 20-years maturities are given in Appendix \ref{appendix_numerical}.


\subsubsection{Market consistency analysis}

The calibration of the DD-SV-LMM is performed on market swaption volatilities for the three following methods: the Heston method, the Edgeworth expansion applied to prices (Proposition \ref{prop_swaption}) and to the volatility smile (Proposition \ref{prop_smile}), respectively called Edgeworth pricing and Edgeworth smile in the following graphs.

\begin{figure}[h]
\centering
\includegraphics[scale=0.3]{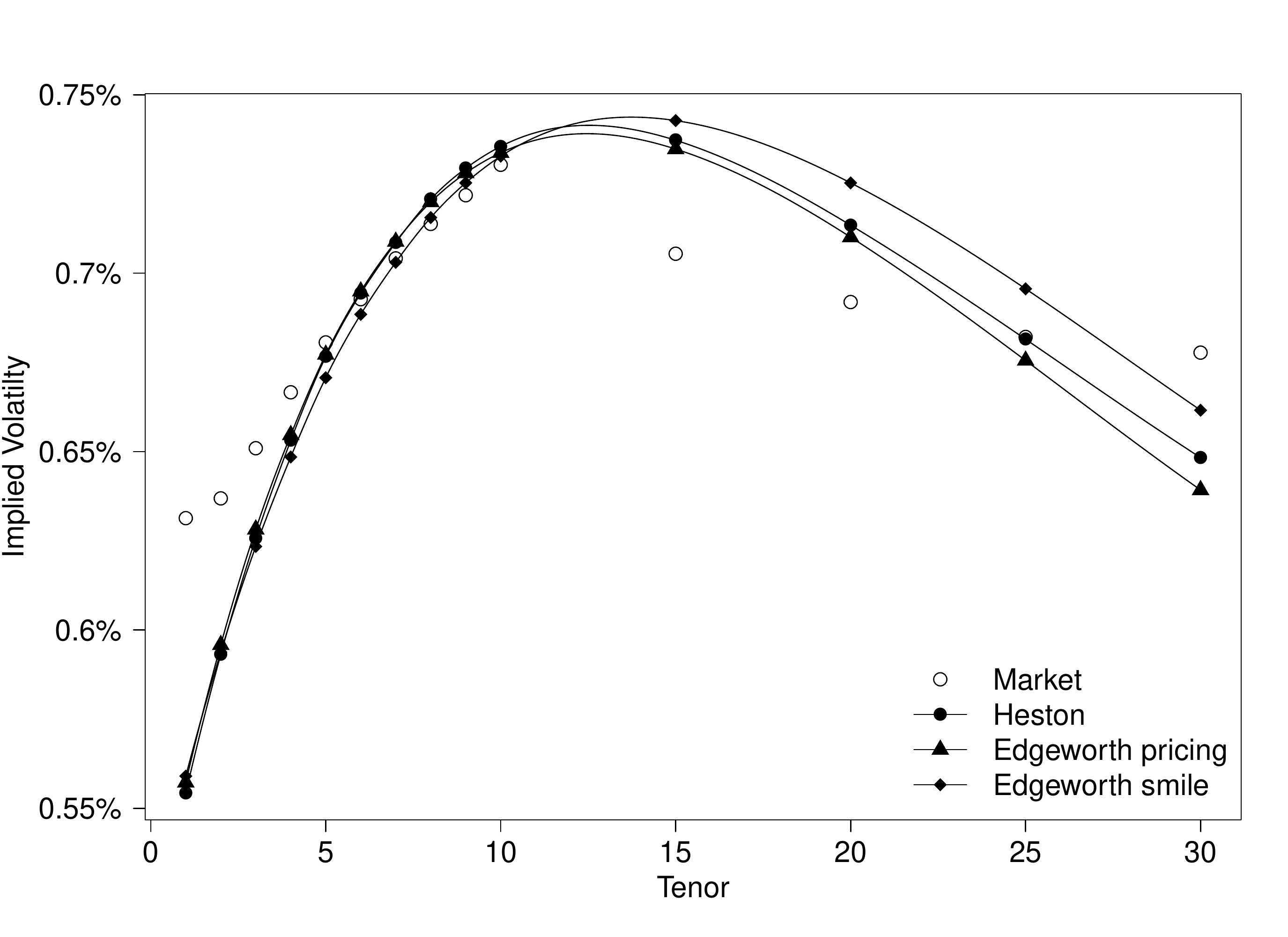}
\caption{ATM Monte Carlo swaption volatilities for 5-years maturity}
\label{figures/graph_5Y_2.pdf}
\end{figure}

We report in Figure \ref{figures/graph_5Y_2.pdf} the ATM empirical swaption volatilities for each method and the corresponding market swaption volatilities for the 5-years maturity. This highlights that the Heston method and both Edgeworth approaches lead to close results, providing a satisfactory Monte Carlo fitting of market data. 

\begin{figure}[h]
\centering
\includegraphics[scale=0.3]{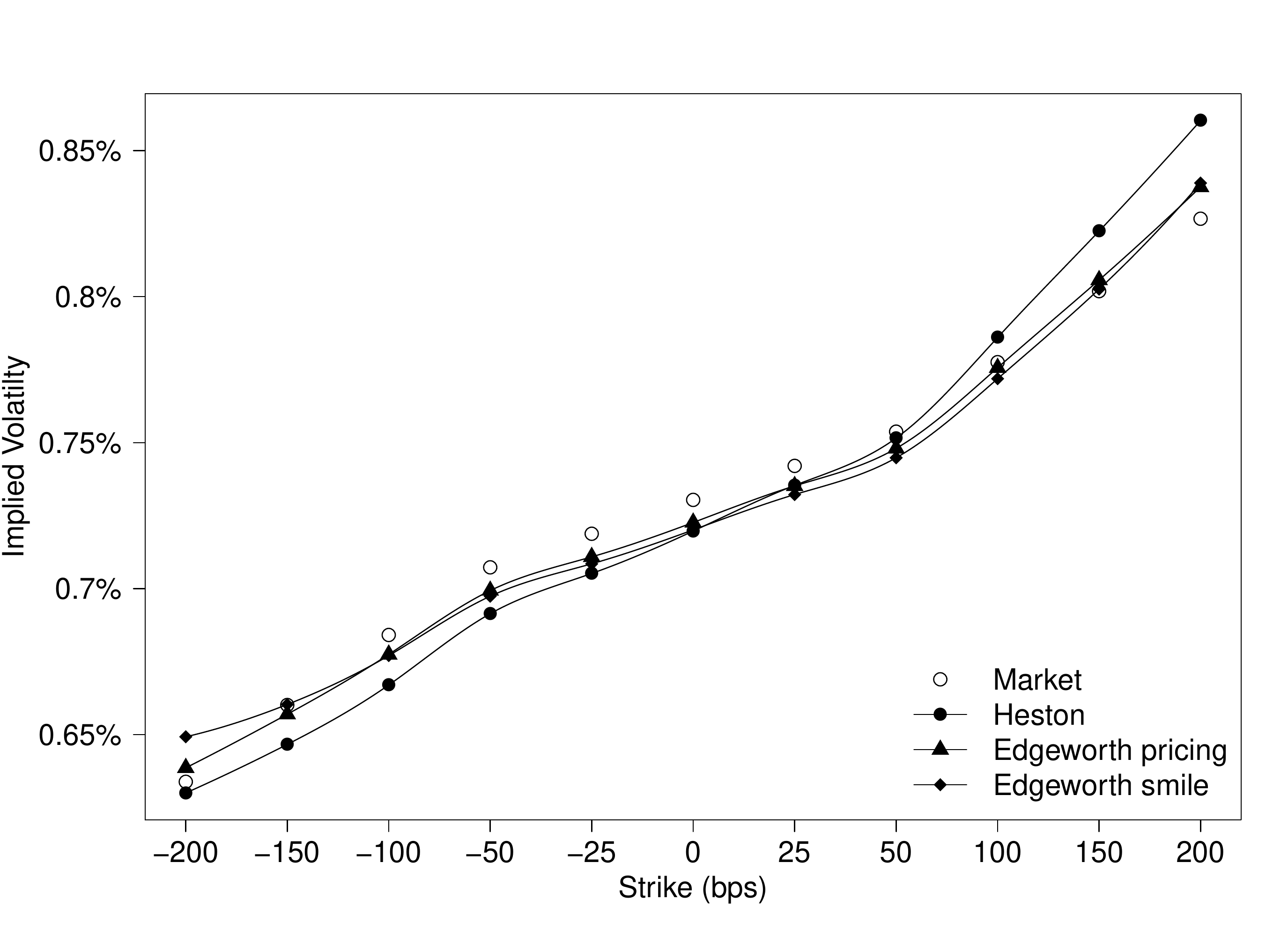}
\caption{Theoretical swaption volatility skews for 5-years maturity}
\label{figures/graph_5Y_3.pdf}
\end{figure}

\begin{figure}[h]
\centering
\includegraphics[scale=0.3]{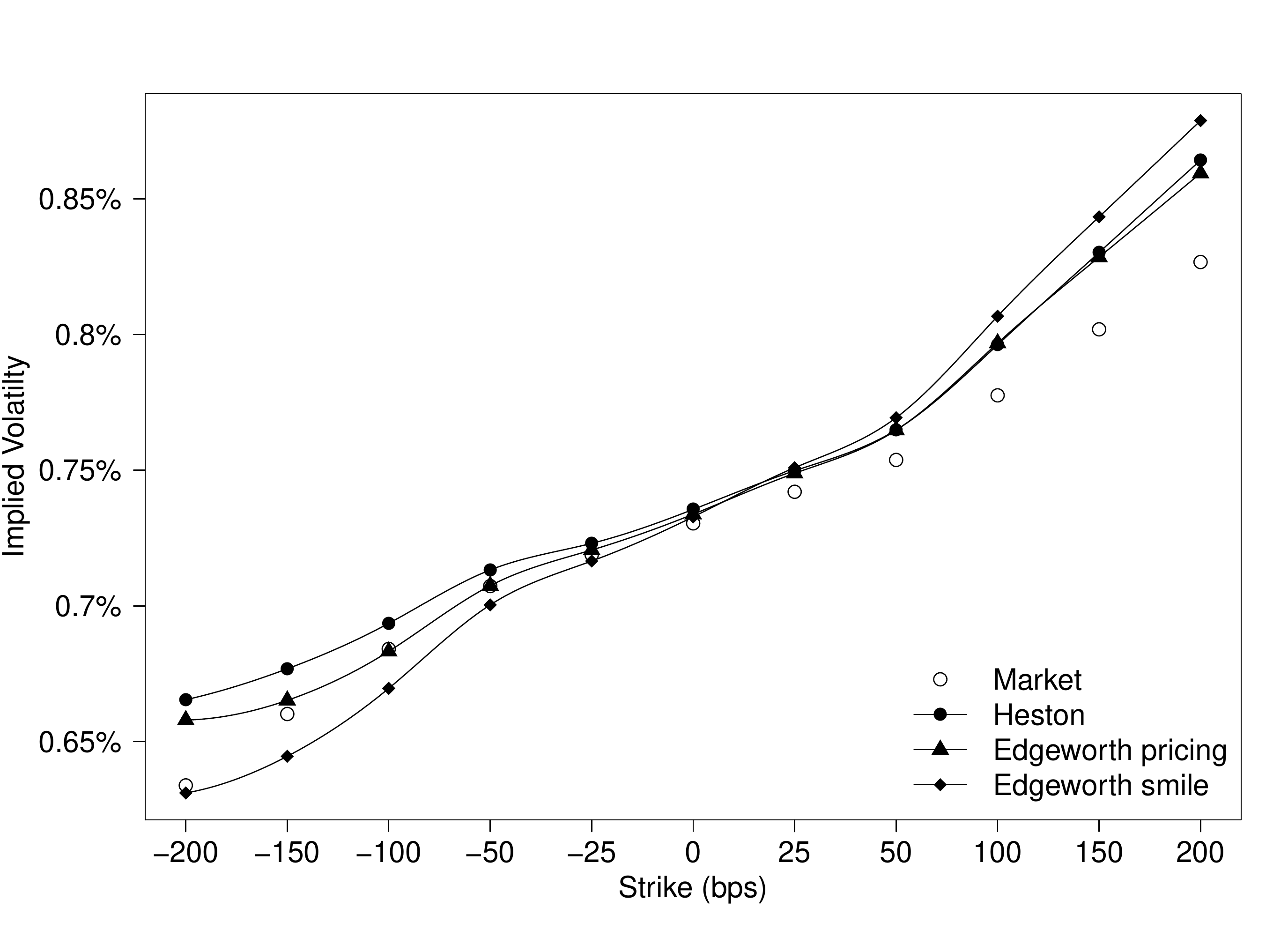}
\caption{Monte Carlo swaption volatility skews for 5-years maturity}
\label{figures/graph_5Y_4.pdf}
\end{figure}

The theoretical (resp. empirical) volatility skew for the 5-years maturity are depicted in Figure \ref{figures/graph_5Y_3.pdf} (resp. Figure \ref{figures/graph_5Y_4.pdf}). It can be seen that the adjustment taken by the Edgeworth expansion implies similar theoretical volatility skews as for the Heston method.  Moreover theoretical and empirical results lead to close volatility skews and strongly support the ability of the Edgeworth approaches to reproduce consistently market quotes. 

Note that the closeness between theoretical and empirical volatilities is more generally discussed in the next section. 


\subsubsection{Accuracy of approximations under a fixed set of parameters}

To assess the accuracy of the approximations underlying each method, we compute for a reference set of parameters theoretical volatilities induced by the Heston method based on numerical integration, and the Edgeworth pricing and  smile formula methods of Propositions \ref{prop_swaption} and \ref{prop_smile}. Then we compare these theoretical elements to Monte Carlo volatilities induced by the reference parameters. We calculate a 95\% confidence interval centered on empirical volatilities and study occurrences of cases where theoretical volatilities are outside confidence intervals.

\begin{figure}[h]
\centering
\includegraphics[scale=0.3]{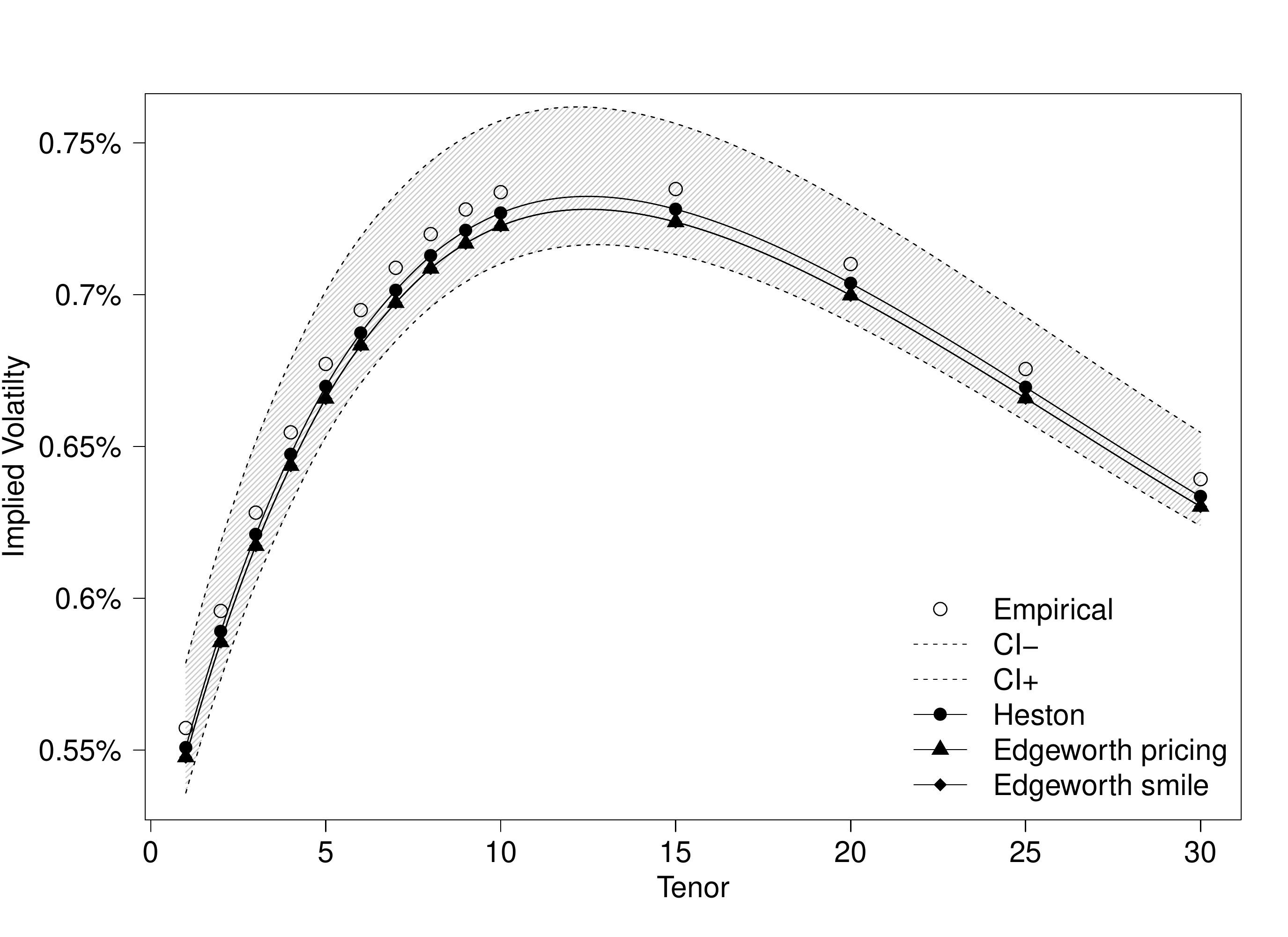}
\caption{ATM swaption volatilities with given parameters for 5-years maturity}
\label{figures/graph_5Y_5.pdf}
\end{figure}

\begin{figure}[h]
\centering
\includegraphics[scale=0.3]{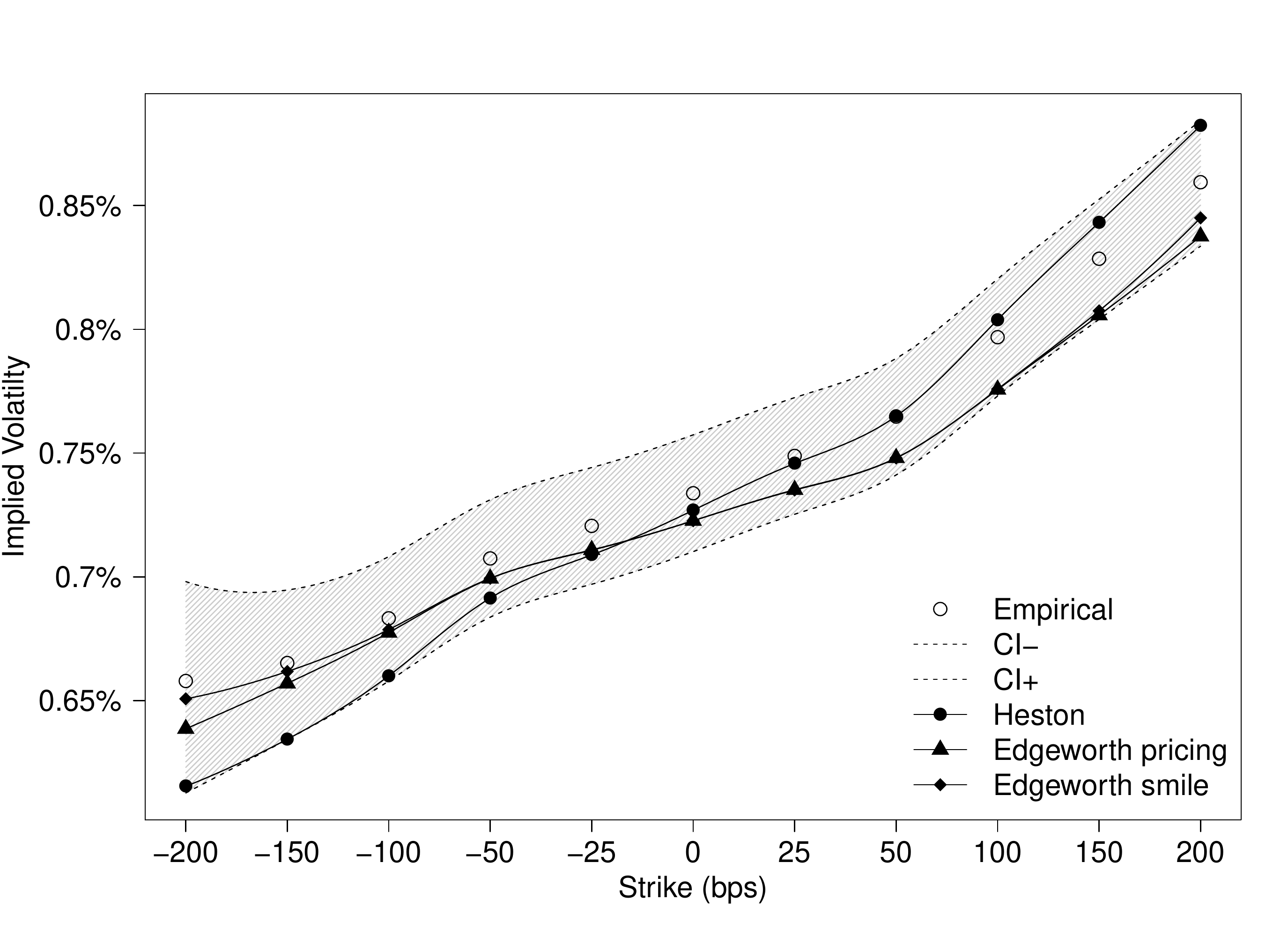}
\caption{Swaption volatility skew with given parameters for 5-years maturity}
\label{figures/graph_5Y_6.pdf}
\end{figure}

For the 5-years maturity, ATM and away-from-the-money theoretical volatilities for all methods lie in the 95\% confidence interval, see Figures \ref{figures/graph_5Y_5.pdf} and \ref{figures/graph_5Y_6.pdf}. This observation supports the robustness of the approximations in both Edgeworth  pricing and smile formulas. 
Differences between theoretical and empirical volatilities find their origin in various reasons. On the one hand, to obtain theoretical volatilities, approximations are taken (freezing technique, and numerical integration for the Heston method, or density approximation for Edgeworth expansion). On the other hand, empirical values are biased by the sampling error, as assessed by the confidence interval, and by the log-Euler discretization scheme used for Monte Carlo simulations.

The Edgeworth pricing and smile methods lead to very similar volatility profiles both for ATM and away-from-the-money swaptions, as shown in Figure \ref{figures/graph_5Y_5.pdf} and \ref{figures/graph_5Y_6.pdf}. 
Based on such analysis, the impact of the approximations involved in the Edgeworth expansion can be assessed; this appears to increase with the maturity, as depicted in Appendix \ref{appendix_numerical} for 10-years and 20-years maturities. Nevertheless, differences between theoretical and Monte Carlo swaption volatilities are small in most cases and strongly back the underlying approximations of Edgeworth expansions.


\subsubsection{Gain in computational time}

Numerical results of this paper have been performed under R 3.2.0, using C++ integration for key functions, in a computer with 2.6 GHz Intel Core i7 CPU. As for the comparison basis, we use a fixed budget of 2500 target function calls in the optimization routine to estimate the parameters of the DD-SV-LMM over 350 swaption volatilities, as detailed in Subsection \ref{subsection_calibration_setting}.
We report in Table \ref{Table_CPU} the CPU time in seconds needed for the calibration of the DD-SV-LMM, on a common basis of 2500 iterations budget for the optimization routine.
The Edgeworth method appears much faster as it provides a 98\% reduction in computational time compared to the classical Heston method.

\begin{table}[h]
\begin{center}
\begin{tabular}{|c | c |}
    \hline
      Method & CPU Time (seconds) \tabularnewline
    \hline
    Heston & 425.1  \tabularnewline%
	\hline
      Edgeworth &  8.2  \tabularnewline
	\hline
 \end{tabular}
 \end{center}\vspace{-0.5cm}
       \caption{CPU time required for calibration using a 2500 optimization iterations budget}
       \label{Table_CPU}
 \end{table}

This result can be mainly explained by the fact that the Heston method involves numerical integration and requires to work in the complex field, whereas the Edgeworth expansion approach takes advantage of the analytical form of swap rate moments up to fourth order, without any numerical differentiation. Furthermore, in the Edgeworth case, the derivatives of the moment generating function are only evaluated for $z=0$, leading to simplified calculations. 

Finally, note that a calibration consisting in minimizing differences between market volatilities and those given by the Edgeworth smile formula is even simpler than the Edgeworth pricing method,  as it doesn’t require numerical inversion of the Bachelier formula during the calibration process.

The gain in speed with Edgeworth expansion enables fast recalibrations of the DD-SV-LMM and can be useful in a variety of topics faced by insurance companies, such as the computation of the Solvency Capital Requirement through Nested Simulations, see \cite{DEVINEAU2009} and \cite{BAUER2012}, the implementation of intensive recalibration process within a Least Squares Monte Carlo framework, see \cite{DEVINEAU2013}, as well as for Variable Annuities hedging and the computation of trading grids. As a matter of fact, the necessity of multiple repeated calibrations for stress-test scenarios involves a rising need for faster calibration processes. For this reason, the Edgeworth pricing and the related smile formula seem to be particularly efficient methods in an operational context.

\section*{Concluding remarks}
In this paper, we illustrated the efficiency of using Edgeworth and Gram-Charlier expansions applied to the calibration of the Libor Market Model with  Stochastic Volatility and Displaced Diffusion (DD-SV-LMM). Our approach brings together two research areas; first, the results regarding the SV-LMM since the work of  \cite{WU2006}, especially on the moment generating function, and second the approximation of density distributions based on Edgeworth or Gram-Charlier expansions. By exploring the analytical tractability of moments up to fourth order, we are able to perform an adjustment of the reference Bachelier model with normal volatilities for skewness and kurtosis, and as a by-product to derive a smile formula relating the volatility to the moneyness with interpretable parameters. The numerical results illustrated in this paper strongly back the approximations involved in the Edgeworth expansion methods, while providing satisfactory results for the fitting of market swaption volatilities. As a main conclusion, our numerical results show a 98\% reduction in computational time in the DD-SV-LMM calibration process compared to the classical Heston method. It is worth mentioning again that our method works on the set of real numbers, making it much more simple and stable compared to the classical Heston approach using numerical integration in the complex field.
As for further research, our method can be extended to any (even) order beyond four, so as to refine the fitting accuracy while keeping the advantage of an efficient computational approach. This could be achieved by the computation of more analytical derivatives, and a deeper understanding of the definition domain of higher order polynomials.

\newpage

\section{Appendices}

\subsection{Solving the moment generating function}
\label{appendix_solving}
Let us consider the separable form of the moment generating function introduced in in Equation (\ref{equation_moment_generating_function}): 
$
	\psi\left( x,V,t;z \right)=e^{A(\tau,z)+B(\tau,z)V+zx}
$
, with $\tau=T_m-t$. Then the first order derivatives of $\psi$ can be computed as
\begin{equation*}
		\frac{\partial \psi}{\partial t} =\left[ -\frac{\partial A}{\partial \tau}-V\frac{\partial B}{\partial \tau} \right]\psi, \; \frac{\partial \psi}{\partial V} =B\psi, \; \frac{\partial \psi}{\partial x} =z\psi,
\end{equation*}
and second order derivatives as
\begin{equation*}
		\frac{\partial^2 \psi}{{\partial V}^2}=B^2\psi, \; \frac{\partial^2 \psi}{{\partial x}^2}=z^2\psi, \; \frac{\partial^2 \psi}{\partial V \partial x}=zB\psi.
\end{equation*}
The Kolmogorov equation in (\ref{equation_kolmogorov}) then becomes
\begin{equation*}
	\left[ -\frac{\partial A}{\partial \tau} +\kappa \theta B\right]+V\left[ -\frac{\partial B}{\partial \tau}-\kappa \xi B +\frac{1}{2}\epsilon^2 B^2 +\epsilon\rho\lambda z B+\frac{1}{2}\lambda^2 z^2 \right]=0.
\end{equation*}
By identification, this leads to the partial differential equation in (\ref{equation_partial}).

\subsection{On Edgeworth expansions}
\label{appendix_edgeworth}

We consider in this appendix the notations introduced in Section \ref{section_edgeworth}.
Let us recall that Edgeworth expansions are used to approximate the cumulative distribution function of a standardized sum of random variables as
\begin{equation}\label{edg_ineq}
		\P(S_n\le x) \approx
			 \Phi(x)
			 - \frac{\gamma_3}{6\sqrt{n}}\varphi(x)H_2(x)
			 + \frac{\gamma_4-3}{24n}\varphi(x)H_3(x)
			 + \frac{\gamma_3^2}{72n}\varphi(x)H_5(x).
\end{equation}
Note that the term $n$ does not appear in papers which aim to derive pricing closed-form expressions. This issue is discussed in \cite{BALIEIRO2004}, where the authors indicate that the term $n$ is incorporated to skewness and kurtosis coefficients, but these considerations are left to the reader.
We propose here to further detail those aspects. Denoting $\mu_3=\E\left[S_n^3  \right]$ and $\mu_4=\E\left[S_n^4  \right]$, the skewness and kurtosis of $S_n$, one recovers that $\mu_3=\frac{\gamma_3}{\sqrt n}$ and $\mu_4=\frac{\gamma_4+3(n-1) }{n}$.
Hence Equation (\ref{edg_ineq}) may be rewritten:
\begin{equation*}
	\P(S_n\le x) \approx
		 \Phi(x)
		 - \frac{\mu_3}{6}\varphi(x)H_2(x)
		 + \frac{\mu_4-3}{24}\varphi(x)H_3(x)
		 + \frac{\mu_3^2}{72}\varphi(x)H_5(x).
\end{equation*}
This corresponds for instance to the formula used for Edgeworth Pricing adjustments in \cite{BALIEIRO2004}.
In the framework developed in Section \ref{section_edgeworth}, we apply this expansion to the standardized variable $Z$ of the swap rate defined in Equation (\ref{equation_Z}):
$Z=\frac
		{R_{m,n}(T_m)-R_{m,n}(0)}
		{\nu}. 
$ 
This random variable is assumed to be (finitely) divisible, that is there exists a (possibly large) integer $n$ and a collection of i.i.d. random increments $(X_j)$ such that
\begin{equation*}
	Z=S_n=\frac{1}{\sqrt n}\sum\limits_{j=1}^n X_j.
\end{equation*}
Note that this includes the set of infinitely divisible distributions, for which the previous decomposition holds for any $n$, as well as stable distributions which are special cases of infinitely divisible ones.
As the calculation of skewness and kurtosis based on moment generating function focuses directly on the variable $Z$, consequently the coefficients $\mu_3$ and $\mu_4$ are homogeneous to those considered in our Edgeworth expansion. For this reason we omit the term $n$ in our Edgeworth framework.

\subsection{Moments for the swap rate distribution}
\label{appendix_moments}

Let us denote $h^{(k)}(z)=\frac{\partial ^k h}{{\partial z}^k} (z)$ for any function $h$, with $h^{(0)}(z)=h(z)$, and write
\begin{equation*}
 	\begin{split}
		A_j^{(0)}(z)&=A\left(\tau_j,z\right),\\
		B_j^{(0)}(z)&=B\left(\tau_j,z\right).
	\end{split}
\end{equation*}
We recall that, denoting $V=V\left(0\right)$ and $\psi^{(0)}(z)=\psi\left(R_{m,n}(0),V,0;z \right)$, the moment generating function writes
\begin{equation*}
	\psi^{(0)}(z)=e^{A^{(0)}_m(z)+B^{(0)}_m(z) V+zR_{m,n}(0)}.
\end{equation*}
Let us define the following functions of $z$:
 \begin{equation*}
 	\begin{split}
		a^{(0)}&=\kappa \xi-\rho \epsilon \lambda z,\\
		d^{(0)}&=\sqrt{
 			\left(a^{(0)}\right)^2 - \lambda^2 \epsilon^2 z^2
 		},\\
 		g_j&=\frac{a^{(0)}+d^{(0)}-\epsilon^2 B_j^{(0)}}{a^{(0)}-d^{(0)}-\epsilon^2 B_j^{(0)}}.
	\end{split}
\end{equation*}
Let us denote $u=\tau_{j+1}-\tau$ and
 \begin{equation*}
 	\begin{split}
	h_1^{(0)}&=a^{(0)}+d^{(0)}-\epsilon^2 B_j^{(0)}, \\ 
	h_2^{(0)}&=1-\exp\left( d^{(0)}u\right), \\ 
	h_3^{(0)}&=a^{(0)}-d^{(0)}-\epsilon^2 B_j^{(0)}, \\ 
	h_4^{(0)}&=a^{(0)}-d^{(0)}-\epsilon^2 B_j^{(0)}-\left( a^{(0)}+d^{(0)}-\epsilon^2 B_j^{(0)} \right)\exp\left( d^{(0)}u\right) \\
	&=h_3^{(0)}+h_1^{(0)}\left(h_2^{(0)}-1\right), \\  
	h_5^{(0)}&=\left(a^{(0)}+d^{(0)}-\epsilon^2 B_j^{(0)} \right)\left(1-\exp\left( d^{(0)}u\right) \right)\left(a^{(0)}-d^{(0)}-\epsilon^2 B_j^{(0)} \right), \\ 
	&=h_1^{(0)}h_2^{(0)}h_3^{(0)}. \nonumber 
	\end{split}
\end{equation*}
Functions $a$, $d$, $g_j$ and $(h_k)$ are here implicitly time dependent
; the recursive scheme is given by
\begin{equation*}
	\left\{
		\begin{split}
		 	A^{(0)}(\tau,z)&=A_j^{(0)}(z)+\tA_j^{(0)}(\tau,z),~~~ \forall  \tau \in (\tau_j,\tau_{j+1}], \\
			B^{(0)}(\tau,z)&=B_j^{(0)}(z)+\tB_j^{(0)}(\tau,z),~~~\forall  \tau \in (\tau_j,\tau_{j+1}], 
		\end{split}
	\right.
\end{equation*}
with
\begin{equation*}
		\begin{split}
		 	\tA_j^{(0)}&=
		 		\frac{\kappa \theta}{\epsilon^2} \left[\left(a^{(0)}+d^{(0)}\right)u-2 \ln \left(\frac{1-g_j \exp\left( d^{(0)}u\right)}{1-g_j }\right) \right], \\
		 		&=\frac{\kappa \theta}{\epsilon^2} \left[\left(a^{(0)}+d^{(0)}\right)u-2 \ln \left(
		 			-\frac
		 				{a^{(0)}-d^{(0)}-\epsilon^2 B_j^{(0)}-\left( a^{(0)}+d^{(0)}-\epsilon^2 B_j^{(0)} \right)\exp\left( d^{(0)}u\right)}
		 				{2d^{(0)}}
		 		\right)\right], \\
		 		&=\frac{\kappa \theta}{\epsilon^2} \left[\left(a^{(0)}+d^{(0)}\right)u-2 \ln \left(
		 			-\frac
		 				{h_4^{(0)}}
		 				{2d^{(0)}}
		 		\right)\right], \\
			\tB_j^{(0)}&=
				\frac{1}{\epsilon^2}
				\frac
					{\left(a^{(0)}+d^{(0)}-\epsilon^2 B_j^{(0)}\right)\left(1-\exp\left( d^{(0)}u\right)\right)}
					{1-g_j \exp\left( d^{(0)}u\right)}, \\
				&=
				\frac{1}{\epsilon^2}
				\frac
					{\left(a^{(0)}+d^{(0)}-\epsilon^2 B_j^{(0)}\right)\left(1-\exp\left( d^{(0)}u\right)\right)\left(a^{(0)}-d^{(0)}-\epsilon^2 B_j^{(0)}\right)}
					{\left(a^{(0)}-d^{(0)}-\epsilon^2 B_j^{(0)}\right)-\left(a^{(0)}+d^{(0)}-\epsilon^2 B_j^{(0)}\right)\exp\left( d^{(0)}u\right)}, \\
				&=
				\frac{1}{\epsilon^2}
				\frac
					{h_5^{(0)}}
					{h_4^{(0)}}. 
		\end{split}
\end{equation*}
 Note that solely the term 
		$d^{(0)}=\sqrt{\left(a^{(0)}\right)^2 - \lambda^2 \epsilon^2 z^2}$
	is different to the one considered in \cite{WU2006}.
As the state variable is $R_{m,n}(t)$, this leads to the additionnal $R_{m,n}(0)z$ term in the moment generating function. 

\subsubsection*{Order 1 derivative}
The first derivative of the moment generating function writes
\begin{equation*}
	\psi^{(1)}=\left( A^{(1)}_m+B^{(1)}_m V +R_{m,n}(0)\right)\psi^{(0)},
\end{equation*}
where the recursive scheme for $j =0,...,m-1$ is given by
\begin{equation*}
	\left\{
		\begin{split}
		 	A^{(1)}(\tau,z)&=A_j^{(1)}(z)+\tA_j^{(1)}(\tau,z),~~~ \forall  \tau \in (\tau_j,\tau_{j+1}], \\
			B^{(1)}(\tau,z)&=B_j^{(1)}(z)+\tB_j^{(1)}(\tau,z),~~~\forall  \tau \in (\tau_j,\tau_{j+1}] ,
		\end{split}
	\right.
\end{equation*}
\begin{equation*}
	\left\{
		\begin{split}
		 	\tA^{(1)}_j&=\frac{\kappa \theta}{\epsilon^2} 
		 		\left[ \left(a^{(1)}+d^{(1)}\right)u-2\left( \frac{h_4^{(1)}}{h_4^{(0)}}-\frac{d^{(1)}}{d^{(0)}} \right) \right], \\
			\tB^{(1)}_j&=\frac{1}{\epsilon^2}
				\left[ \frac{h_5^{(1)}}{h_4^{(0)}}-\frac{h_5^{(0)}h_4^{(1)}}{\left(h_4^{(0)}\right)^2}  \right], 
		\end{split}
	\right.
\end{equation*}
with
\begin{equation*}
		\begin{split}
			a^{(1)}&=-\rho \epsilon \lambda,\\
			d^{(1)}&=\frac{a^{(0)}a^{(1)}- \lambda^2 \epsilon^2 z}{d^{(0)}},
		\end{split}
\end{equation*}
and
\begin{equation*}
		\begin{split}			
		 	h^{(1)}_1&=a^{(1)}+d^{(1)}-\epsilon^2 B^{(1)}_j, \\ 
			h^{(1)}_2&=-d^{(1)}u\exp\left( d^{(0)}u\right), \\ 
			h^{(1)}_3&=a^{(1)}-d^{(1)}-\epsilon^2 B^{(1)}_j,\\ 
			h^{(1)}_4&=h^{(1)}_3+h^{(1)}_1\left(h_2^{(0)}-1\right)+h_1^{(0)}h^{(1)}_2,\\  
			h^{(1)}_5&=h^{(1)}_1h_2^{(0)}h_3^{(0)}+h_1^{(0)}h^{(1)}_2h_3^{(0)}+h_1^{(0)}h_2^{(0)}h^{(1)}_3.
		\end{split}
\end{equation*}

\subsubsection*{Order 2 derivative}
The second derivative of the moment generating function writes
\begin{equation*}
	\psi^{(2)} =
		\left( A_m^{(2)}+B_m^{(2)}V\right) \psi^{(0)}
		+\frac{{( \psi^{(1)})}^2}{\psi^{(0)}},
\end{equation*}
where the recursive scheme for $j =0,...,m-1$ is given by
\begin{equation*}
	\left\{
		\begin{split}
		 	A^{(2)}(\tau,z)&=A_j^{(2)}(z)+\tA_j^{(2)}(\tau,z),~~~ \forall  \tau \in (\tau_j,\tau_{j+1}], \\
			B^{(2)}(\tau,z)&=B_j^{(2)}(z)+\tB_j^{(2)}(\tau,z),~~~\forall  \tau \in (\tau_j,\tau_{j+1}], 
		\end{split}
	\right.
\end{equation*}
\begin{equation*}
	\left\{
		\begin{split}
		 	\tA^{(2)}_j&=\frac{\kappa \theta}{\epsilon^2} 
		 		\left[d^{(2)}u-2\left(
		 			 \frac{h^{(2)}_4}{h_4^{(0)}}
		 			 -\frac{d^{(2)}}{d^{(0)}}
		 			 -\frac{\left(h^{(1)}_4\right)^2}{\left(h_4^{(0)}\right)^2}
		 			 +\frac{\left(d^{(1)}\right)^2}{\left(d^{(0)}\right)^2} 
				\right) \right], \\
			\tB^{(2)}_j&=\frac{1}{\epsilon^2}
				\left[ 
					\frac{h^{(2)}_5}{h_4^{(0)}}
					-\frac{h_5^{(0)}h_4^{(2)}}{\left(h_4^{(0)}\right)^2}
					-\frac{2h^{(1)}_5h_4^{(1)}}{\left(h_4^{(0)}\right)^2}
					+\frac{2h_5^{(0)}\left(h_4^{(1)}\right)^2}{\left(h_4^{(0)}\right)^3}
				\right],
		\end{split}
	\right.
\end{equation*}
with
\begin{equation*}
		d^{(2)}=\frac{\left(a^{(1)}\right)^2-\lambda^2\epsilon^2-\left(d^{(1)}\right)^2}{d^{(0)}},
\end{equation*}
and since $a^{(2)} =0$,
\begin{equation*}
	\begin{split}
	 	h^{(2)}_1&=d^{(2)}-\epsilon^2 B^{(2)}_j, \\ 
		h^{(2)}_2&=-d^{(2)}u\exp\left( d^{(0)}u\right)+d^{(1)}uh^{(1)}_2, \\ 
		h^{(2)}_3&=-d^{(2)}-\epsilon^2 B^{(2)}_j,\\ 
		h^{(2)}_4&=
			h^{(2)}_3
			+h^{(2)}_1\left(h_2^{(0)}-1\right)
			+h_1^{(0)}h_2^{(2)}
			+2h^{(1)}_1h^{(1)}_2,\\  
		h^{(2)}_5&=
			h^{(2)}_1h_2^{(0)}h_3^{(0)}
			+h_1^{(0)}h^{(2)}_2h_3^{(0)}
			+h_1^{(0)}h_2^{(0)}h^{(2)}_3
			+2\left(
				h^{(1)}_1h^{(1)}_2h_3^{(0)}
				+h_1^{(1)}h^{(0)}_2h^{(1)}_3
				+h_1^{(0)}h^{(1)}_2h^{(1)}_3
			\right).
	\end{split}
\end{equation*}

\subsubsection*{Order 3 derivative}
The third derivative of the moment generating function writes
\begin{equation*}
	\psi^{(3)}=
 	\left(  A_m^{(3)}+B_m^{(3)}V\right)\psi^{(0)}
 	+\left(  A_m^{(2)}+B_m^{(2)}V\right)\psi^{(1)}
 	+ \frac{2\psi^{(2)} \psi^{(1)}}{\psi^{(0)}}
 	-\frac{\left(\psi^{(1)}\right)^3}{\left(\psi^{(0)}\right)^2},
\end{equation*}
where the recursive scheme  for $j =0,...,m-1$ is given by
\begin{equation*}
	\left\{
		\begin{split}
		 	A^{(3)}(\tau,z)&=A_j^{(3)}(z)+\tA_j^{(3)}(\tau,z),~~~ \forall  \tau \in (\tau_j,\tau_{j+1}], \\
			B^{(3)}(\tau,z)&=B_j^{(3)}(z)+\tB_j^{(3)}(\tau,z),~~~\forall  \tau \in (\tau_j,\tau_{j+1}], 
		\end{split}
	\right.
\end{equation*}
\begin{equation*}
	\left\{
		\begin{split}
	 	\tA^{(3)}_j&=\frac{\kappa \theta}{\epsilon^2} 
	 		\left[d^{(3)}u-2
			\left(
	 		 	\frac{h^{(3)}_4}{h_4^{(0)}}
	 		 	-\frac{ d^{(3)} }{ d^{(0)} }
	 		 	-\frac{3h^{(2)}_4 h^{(1)}_4}{\left(h_4^{(0)}\right)^2}
	 		 	+\frac{ 3d^{(2) }d^{(1)} }{ \left(d^{(0)}\right)^2 }
	 		 	+\frac{2\left(h^{(1)}_4\right)^3 }{ \left(h_4^{(0)}\right)^3 }	
	 		 	-\frac{ 2\left(d^{(1)}\right)^3}{ \left(d^{(0)}\right)^3 }
	 		\right) \right], \\
			\tB^{(3)}_j&=\frac{1}{\epsilon^2}
				\left[
					\frac{h^{(3)}_5}{h_4^{(0)}}
					-\frac{ h_5^{(0)}h_4^{(3)} }{ \left(h_4^{(0)}\right)^2 }
					-\frac{ 3h^{(2) }_5h^{(1)}_4 }{ \left(h_4^{(0)}\right)^2 }
					-\frac{3h^{(1)}_5h^{(2)}_4 }{ \left(h_4^{(0)}\right)^2 }
					+\frac{6h_5^{(0)}h^{(2)}_4h^{(1)}_4}{\left(h_4^{(0)}\right)^3}
					+\frac{6h^{(1)}_5\left(h^{(1)}_4\right)^2}{\left(h_4^{(0)}\right)^3}
				\right. \\
				& \left.
					-\frac{6h_5^{(0)}\left(h^{(1)}_4\right)^3}{\left(h_4^{(0)}\right)^4}
				\right], 	
		\end{split}
	\right.
\end{equation*}
with
\begin{equation*}
		d^{(3)} =\frac{-3d^{(2)}d^{(1)}}{d^{(0)}},
\end{equation*}
and
\begin{equation*}
	\begin{split}
	 	h^{(3)}_1&=d^{(3)}-\epsilon^2 B^{(3)}_j, \\ 
		h^{(3)}_2&=
			-d^{(3)}u\exp\left( d^{(0)}u\right)
			+2d^{(2)}uh^{(1)}_2 
			+d^{(1)}uh^{(2)}_2, \\ 
		h^{(3)}_3&=-d^{(3)}-\epsilon^2 B^{(3)}_j,\\ 
		h^{(3)}_4&=
			h^{(3)}_3
			+h^{(3)}_1\left(h_2^{(0)}-1\right)
			+h^{(0)}_1h^{(3)}_2
			+3h^{(2)}_1h^{(1)}_2
			+3h^{(1)}_1h^{(2)}_2,\\  
		h^{(3)}_5&=
			h^{(3)}_1h^{(0)}_2h^{(0)}_3
			+h^{(0)}_1h^{(3)}_2h^{(0)}_3
			+h^{(0)}_1h^{(0)}_2h^{(3)}_3\\
			&+3\left(
				h^{(2)}_1h^{(1)}_2h^{(0)}_3
				+h^{(2)}_1h^{(0)}_2h^{(1)}_3
				+h^{(1)}_1h^{(2)}_2h^{(0)}_3
				+h^{(0)}_1h^{(2)}_2h^{(1)}_3
				+h^{(1)}_1h^{(0)}_2h^{(2)}_3
				+h^{(0)}_1h^{(1)}_2h^{(2)}_3
			\right)\\
			&+6h^{(1)}_1h^{(1)}_2h^{(1)}_3.
	\end{split}
\end{equation*}

\subsubsection*{Order 4 derivative}
The fourth derivative of the moment generating function writes
\begin{equation*}
	\begin{split}
		\psi^{(4)}&=
		 	\left(  A_m^{(4)}+B_m^{(4)}V\right)\psi
		 	+2\left(  A_m^{(3)}+B_m^{(3)}V\right)\psi^{(1)}
		 	+ \left(  A_m^{(2)}+B_m^{(2)}V\right)\psi^{(2)}\\
		 	&+  \frac{2\psi^{(3)} \psi^{(1)}}{\psi^{(0)}}
	 		+ \frac{2 \left(\psi^{(2)}\right)^2}{\psi^{(0)}}
	 		- \frac{5\psi^{(2)}\left(\psi^{(1)}\right)^2 }{\left(\psi^{(0)}\right)^2}
	 		+\frac{2\left(\psi^{(1)}\right)^4}{\left(\psi^{(0)}\right)^3},
	\end{split}
\end{equation*}
where the recursive scheme  for $j =0,...,m-1$ is given by
\begin{equation*}
	\left\{
		\begin{split}
		 	A^{(4)}(\tau,z)&=A_j^{(4)}(z)+\tA_j^{(4)}(\tau,z),~~~ \forall  \tau \in (\tau_j,\tau_{j+1}], \\
			B^{(4)}(\tau,z)&=B_j^{(4)}(z)+\tB_j^{(4)}(\tau,z),~~~\forall  \tau \in (\tau_j,\tau_{j+1}] ,
		\end{split}
	\right.
\end{equation*}
\begin{equation*}
	\left\{
		\begin{split}
		 	\tA^{(4)}_j&=\frac{\kappa \theta}{\epsilon^2} 
		 		\left[d^{(4)}u-2
				\left(
		 		 	\frac{h^{(4)}_4}{h_4^{(0)}}
		 		 	-\frac{ d^{(4)} }{ d^{(0)} }
					-\frac{4h^{(3)}_4h^{(1)}_4}{\left(h_4^{(0)}\right)^2}
					+\frac{4 d^{(3)}d^{(1)} }{ \left(d^{(0)}\right)^2 }
					-\frac{3\left(h^{(2)}_4\right)^2}{\left(h_4^{(0)}\right)^2}
					+\frac{3\left( d^{(2)}\right)^2 }{ \left(d^{(0)}\right)^2 } 
				\right.\right. \\
				&\left.\left.
					+\frac{12h^{(2)}_4\left(h^{(1)}_4\right)^2}{\left(h_4^{(0)}\right)^3}
					-\frac{12 d^{(2)}\left(d^{(1)}\right)^2 }{ \left(d^{(0)}\right)^3 }
					-\frac{6\left(h^{(1)}_4\right)^4}{\left(h_4^{(0)}\right)^4}
					+\frac{6 \left(d^{(1)}\right)^4 }{ \left(d^{(0)}\right)^4 }
		 		\right) \right], \\
			\tB^{(4)}_j&=\frac{1}{\epsilon^2}
				\left[
					\frac{h^{(4)}_5}{h_4^{(0)}}
					-\frac{ h_5^{(0)}h_4^{(4)} }{ \left(h_4^{(0)}\right)^2 }
					-\frac{ 4h^{(3) }_5h^{(1)}_4 }{ \left(h_4^{(0)}\right)^2 }
					-\frac{4h^{(1)}_5h^{(3)}_4 }{ \left(h_4^{(0)}\right)^2 }
					-\frac{6h^{(2)}_5h^{(2)}_4 }{ \left(h_4^{(0)}\right)^2 }
					+\frac{8h^{(0)}_5h^{(3)}_4h^{(1)}_4}{\left(h_4^{(0)}\right)^3}
					+\frac{12h^{(2)}_5\left(h^{(1)}_4\right)^2}{\left(h_4^{(0)}\right)^3}
				\right. \\
				 & \left.
					+\frac{24h^{(1)}_5 h^{(2)}_4 h^{(1)}_4}{\left(h_4^{(0)}\right)^3}
					+\frac{6h^{(0)}_5\left(h^{(2)}_4\right)^2}{\left(h_4^{(0)}\right)^3}
					-\frac{36h^{(0)}_5 h^{(2)}_4\left(h^{(1)}_4\right)^2}{\left(h_4^{(0)}\right)^4}
					-\frac{24h^{(1)}_5\left(h^{(1)}_4\right)^3}{\left(h_4^{(0)}\right)^4}
					+\frac{24h^{(0)}_5 \left(h^{(1)}_4\right)^4}{\left(h_4^{(0)}\right)^5}
				\right], 	
		\end{split}
	\right.
\end{equation*}
with
\begin{equation*}
		d^{(4)}=
			-\frac{3d^{(3)}d^{(1)}}{d^{(0)}}
			-\frac{3\left(d^{(2)}\right)^2}{d^{(0)}}
			+\frac{3d^{(2)}\left(d^{(1)}\right)^2}{\left(d^{(0)}\right)^2},
\end{equation*}
and
\begin{equation*}
	\begin{split}
	 	h^{(4)}_1&=d^{(4)}-\epsilon^2 B^{(4)}_j, \\ 
		h^{(4)}_2&=-d^{(4)}u\exp\left( d^{(0)}u\right)
			+d^{(1)}uh^{(3)}_2 
			+3d^{(3)}uh^{(1)}_2
			+3d^{(2)}uh^{(2)}_2, \\ 
		h^{(4)}_3&=-d^{(4)}-\epsilon^2 B^{(4)}_j,\\ 
		h^{(4)}_4&=h^{(4)}_3
			+h^{(4)}_1\left(h_2^{(0)}-1\right)
			+h_1^{(0)}h_2^{(4)}
			+4h^{(3)}_1h^{(1)}_2
			+4h^{(1)}_1h^{(3)}_2
			+6h^{(2)}_1h^{(2)}_2,\\  
		h^{(4)}_5&=
			h^{(4)}_1h_2^{(0)}h_3^{(0)}
			+h_1^{(0)}h^{(4)}_2h_3^{(0)}
			+h_1^{(0)}h^{(0)}_2h^{(4)}_3\\&+
			4\left(
				h^{(3)}_1h^{(1)}_2h^{(0)}_3
				+h^{(3)}_1h^{(0)}_2h^{(1)}_3
				+h^{(1)}_1h^{(3)}_2h^{(0)}_3
				+h^{(0)}_1h^{(3)}_2h^{(1)}_3
				+h^{(1)}_1h^{(0)}_2h^{(3)}_3
				+h^{(0)}_1h^{(1)}_2h^{(3)}_3
			\right)\\
			&+6\left(
				h_1^{(2)}h_2^{(2)}h^{(0)}_3
				+h_1^{(2)}h^{(0)}_2h_3^{(2)}
				+h^{(0)}_1h_2^{(2)}h_3^{(2)}
			\right)
			+12\left(
				h^{(2)}_1h_2^{(1)}h_3^{(1)}
				+h_1^{(1)}h^{(2)}_2h_3^{(1)}
				+h_1^{(1)}h_2^{(1)}h^{(2)}_3
			\right).
	\end{split}
\end{equation*}

\subsection{Numerical results for maturities 10-years and 20-years}
\label{appendix_numerical}

\begin{figure}[H]
\centering
\includegraphics[scale=0.3]{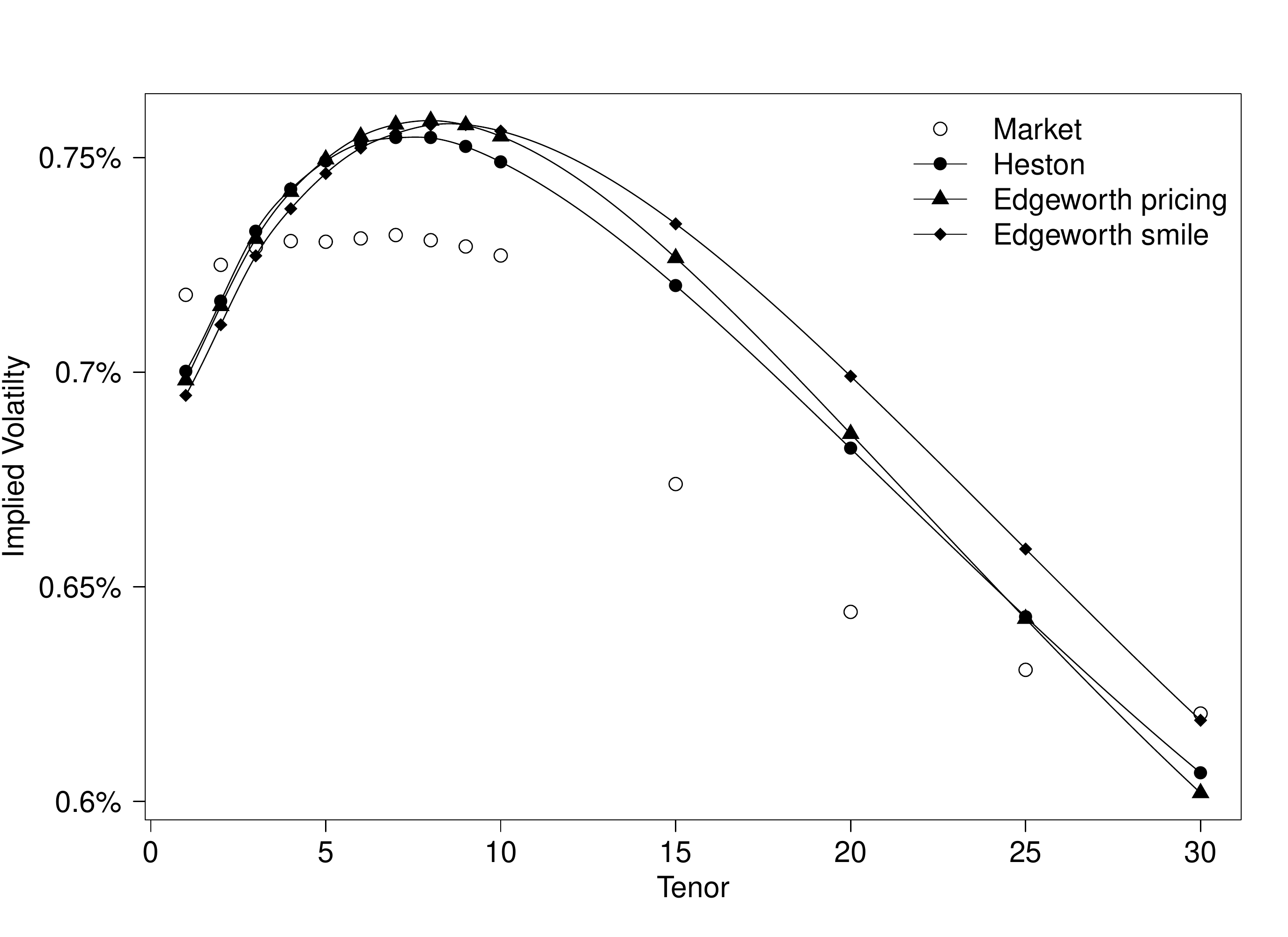}
\caption{ATM Monte Carlo swaption volatilities for 10-years maturity}
\label{figures/graph_10Y_2.pdf}
\end{figure}

\begin{figure}[H]
\centering
\includegraphics[scale=0.3]{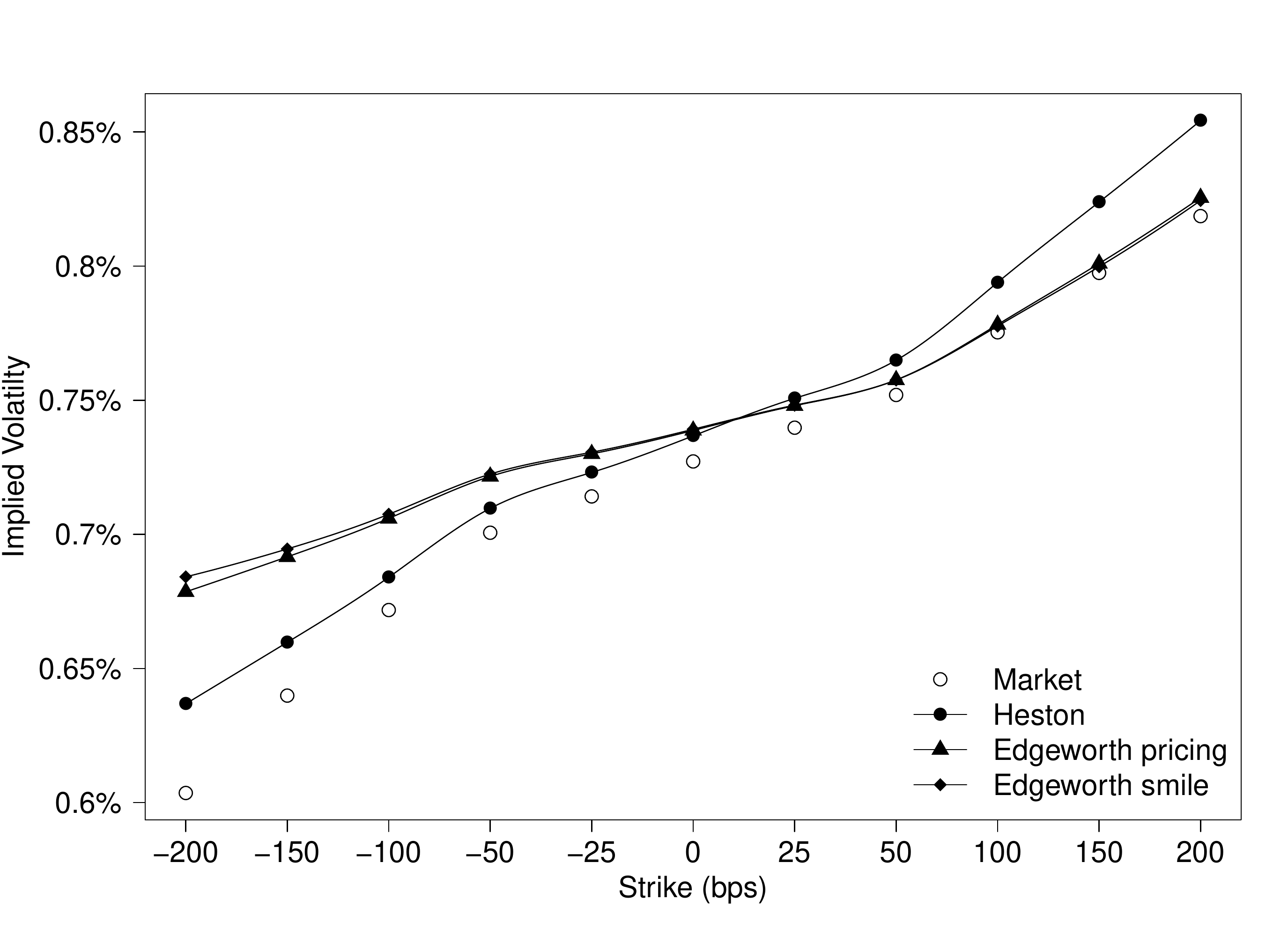}
\caption{Theoretical swaption volatility skews for 10-years maturity}
\label{figures/graph_10Y_3.pdf}
\end{figure}

\begin{figure}[H]
\centering
\includegraphics[scale=0.3]{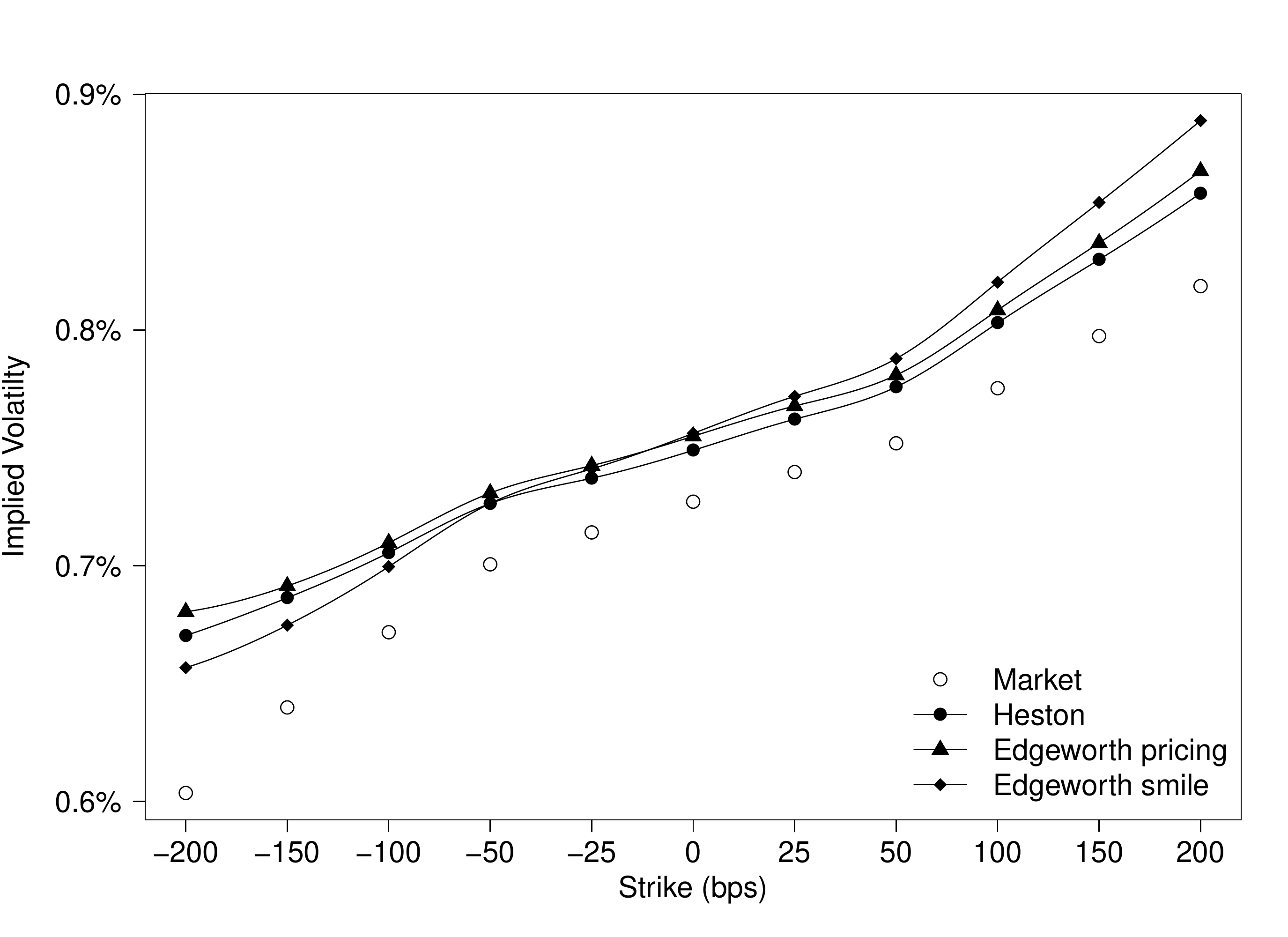}
\caption{Monte Carlo swaption volatility skews for 10-years maturity}
\label{figures/graph_10Y_4.pdf}
\end{figure}

\begin{figure}[H]
\centering
\includegraphics[scale=0.3]{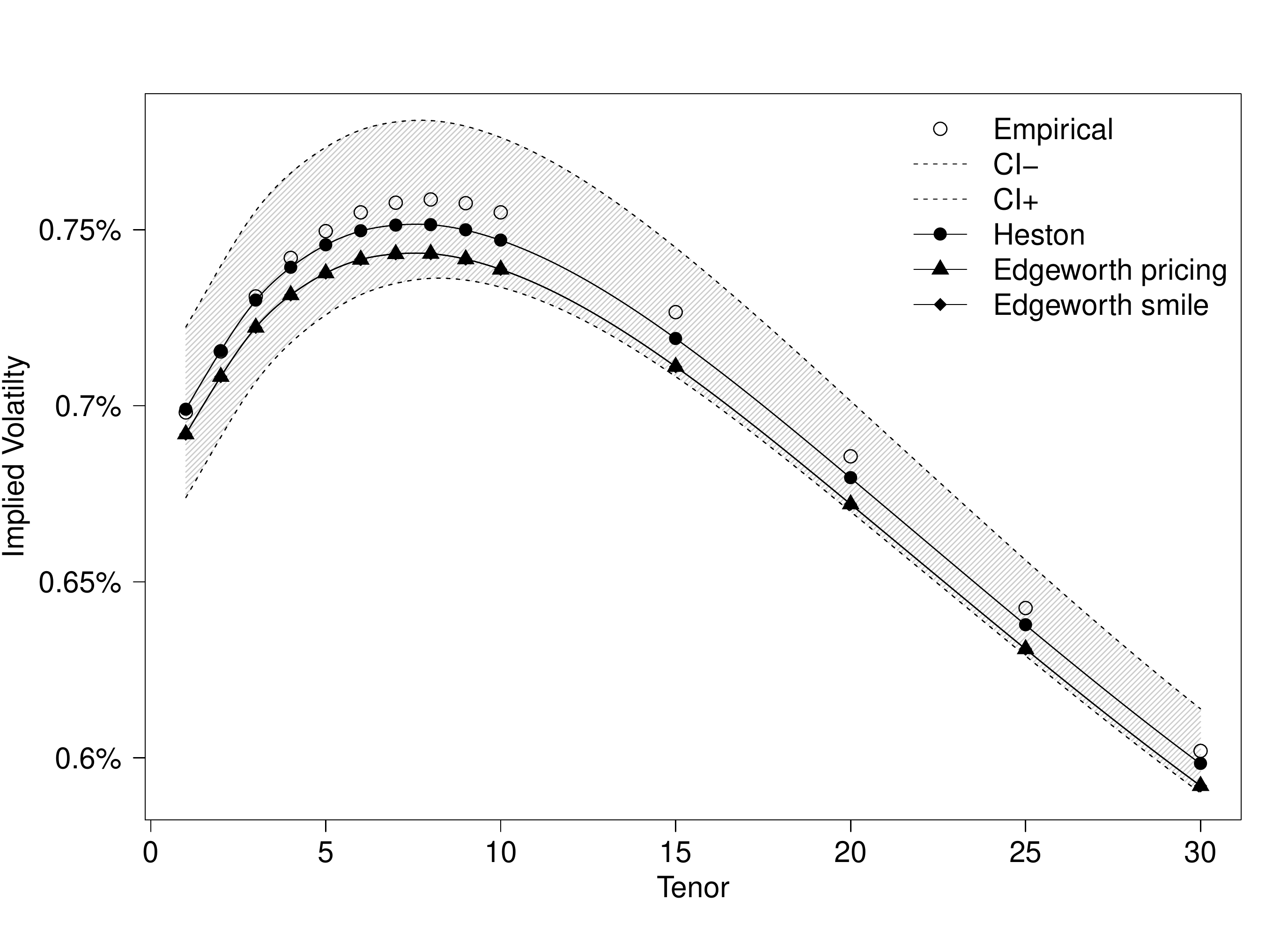}
\caption{ATM swaption volatilities with given parameters for 10-years maturity}
\label{figures/graph_10Y_5.pdf}
\end{figure}

\begin{figure}[H]
\centering
\includegraphics[scale=0.3]{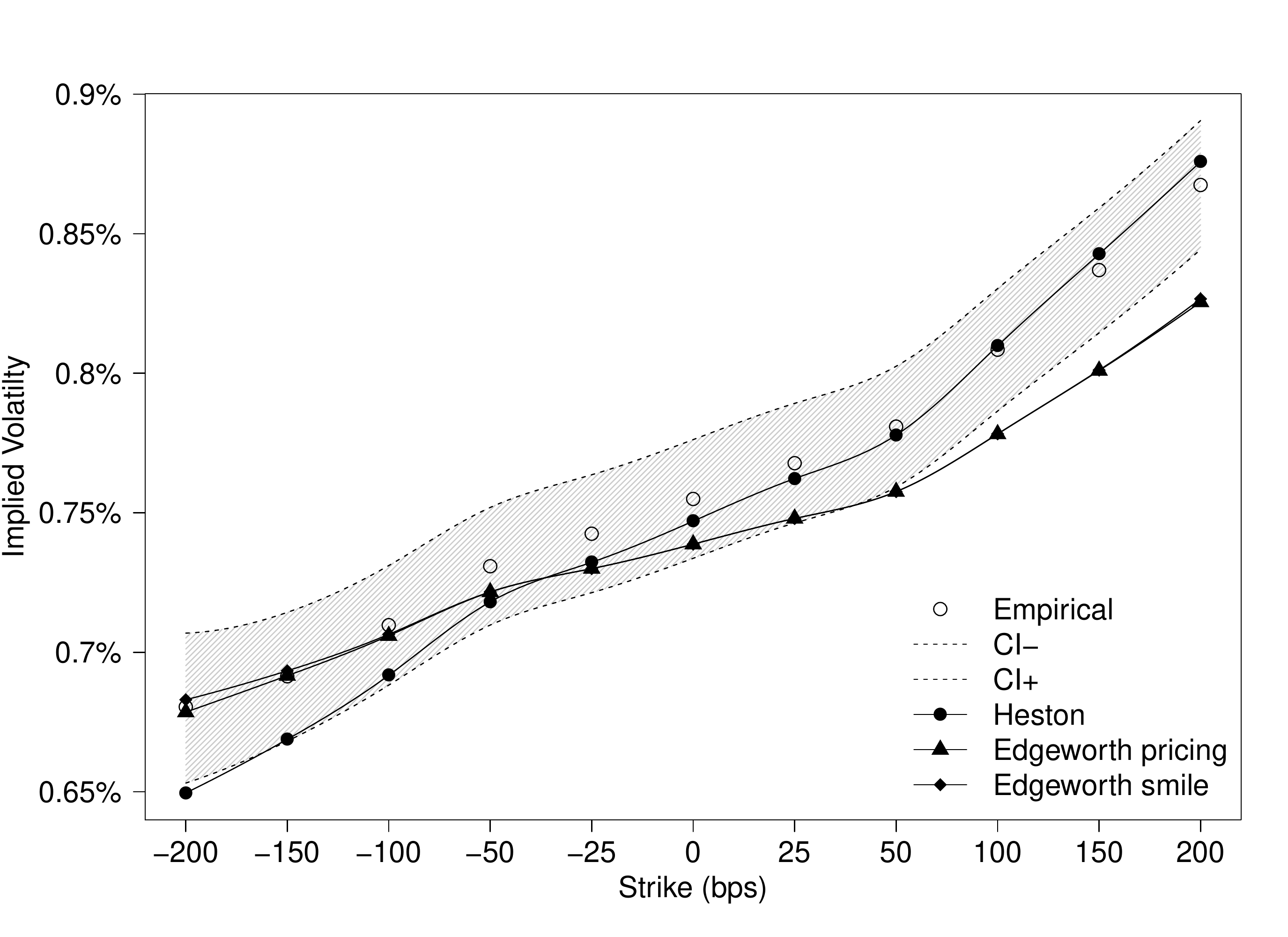}
\caption{Swaption volatility skews with given parameters for 10-years maturity}
\label{figures/graph_10Y_6.pdf}
\end{figure}

\begin{figure}[H]
\centering
\includegraphics[scale=0.3]{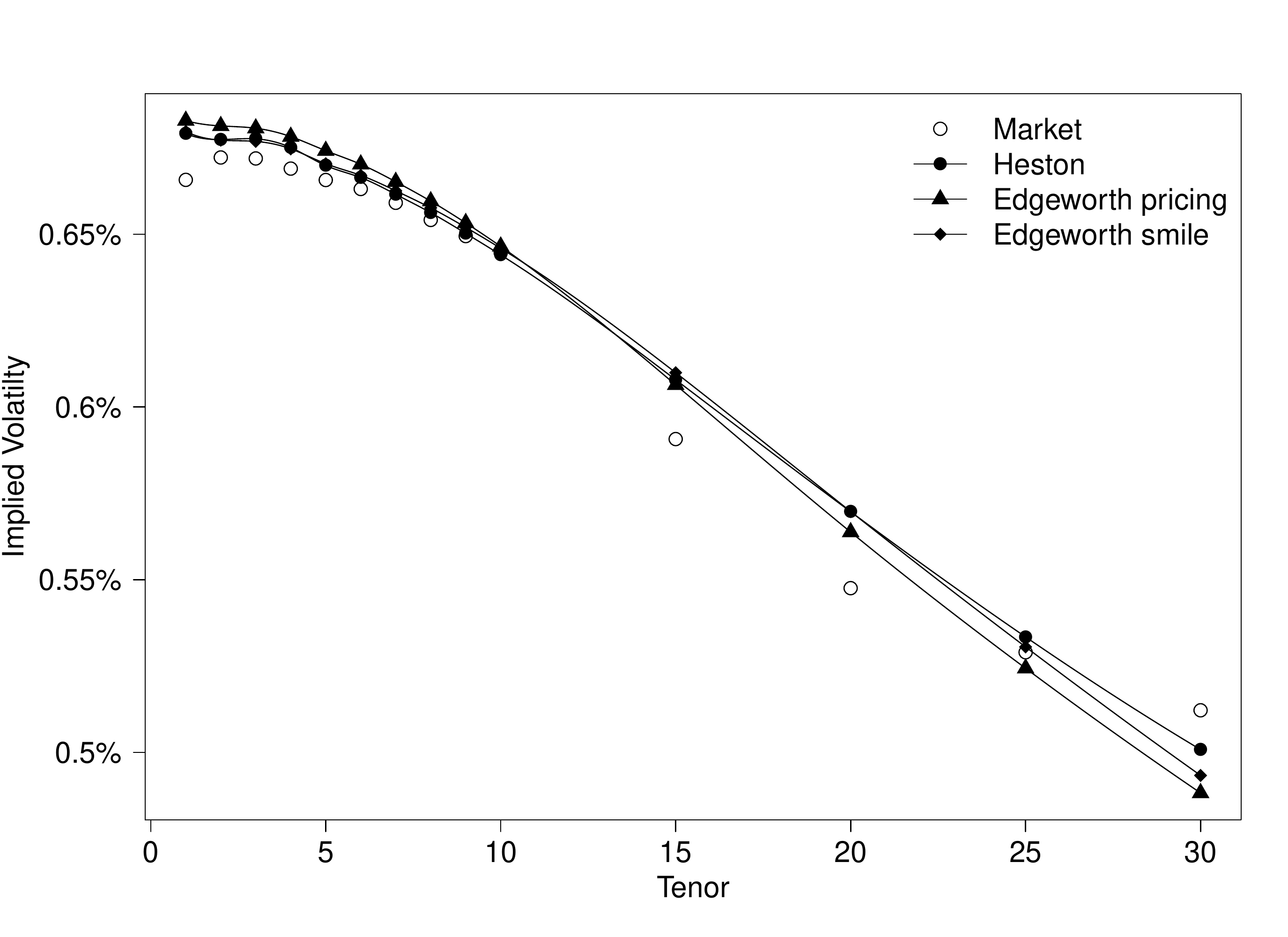}
\caption{ATM Monte Carlo swaption volatilities for 20-years maturity}
\label{figures/graph_20Y_2.pdf}
\end{figure}

\begin{figure}[H]
\centering
\includegraphics[scale=0.3]{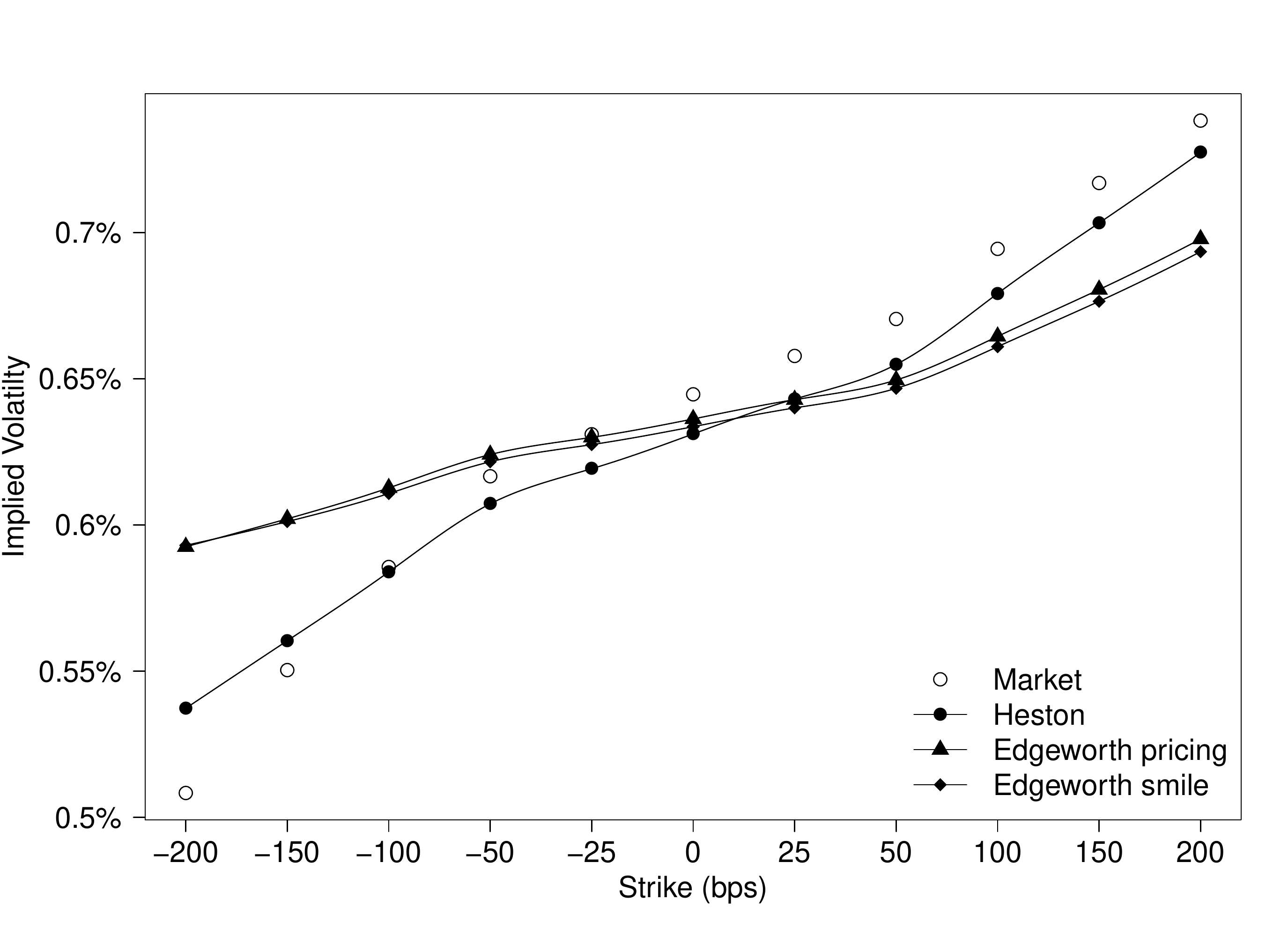}
\caption{Theoretical swaption volatility skews for 20-years maturity}
\label{figures/graph_20Y_3.pdf}
\end{figure}

\begin{figure}[H]
\centering
\includegraphics[scale=0.3]{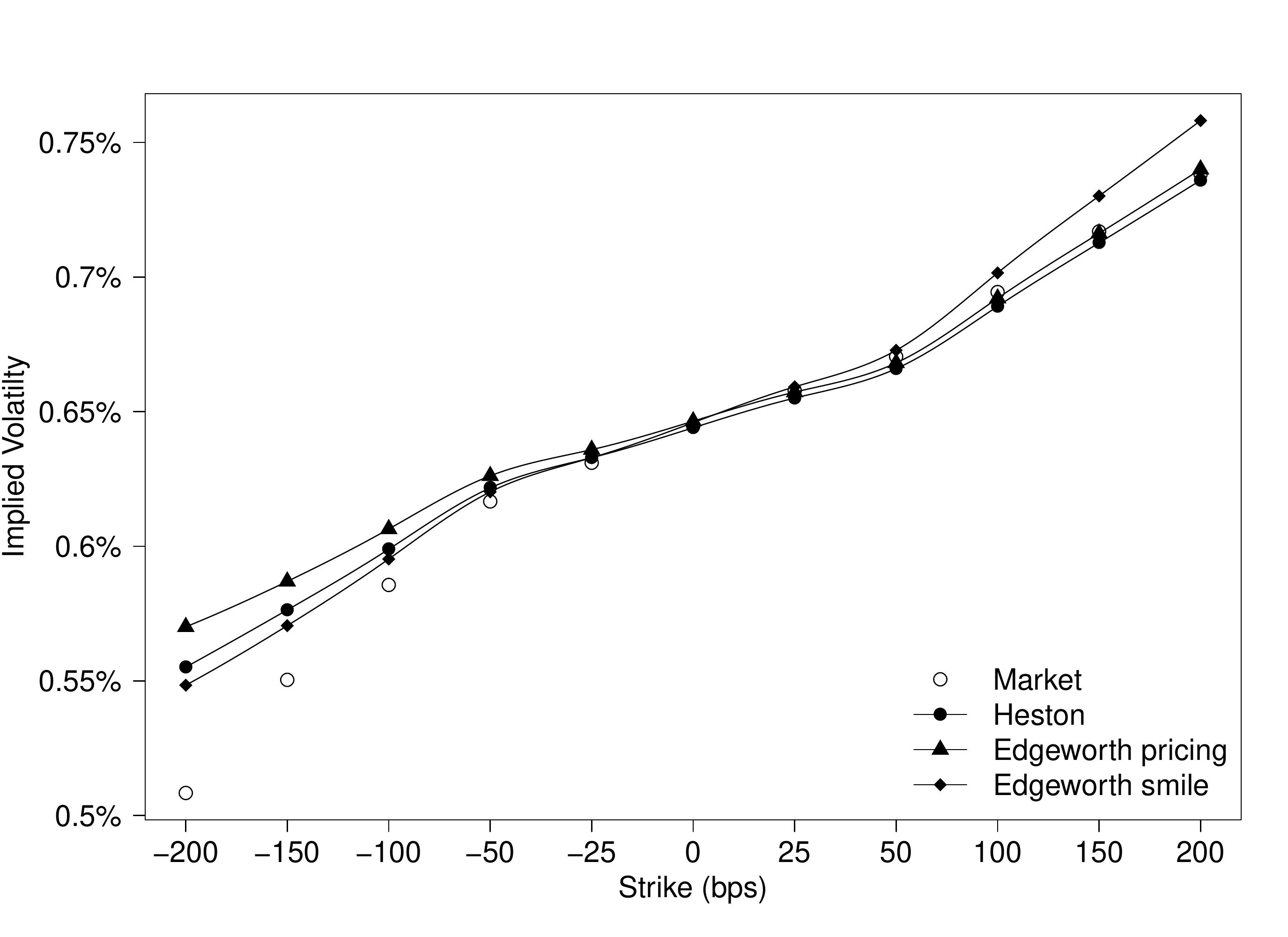}
\caption{Monte Carlo swaption volatility skews for 20-years maturity}
\label{figures/graph_20Y_4.pdf}
\end{figure}

\begin{figure}[H]
\centering
\includegraphics[scale=0.3]{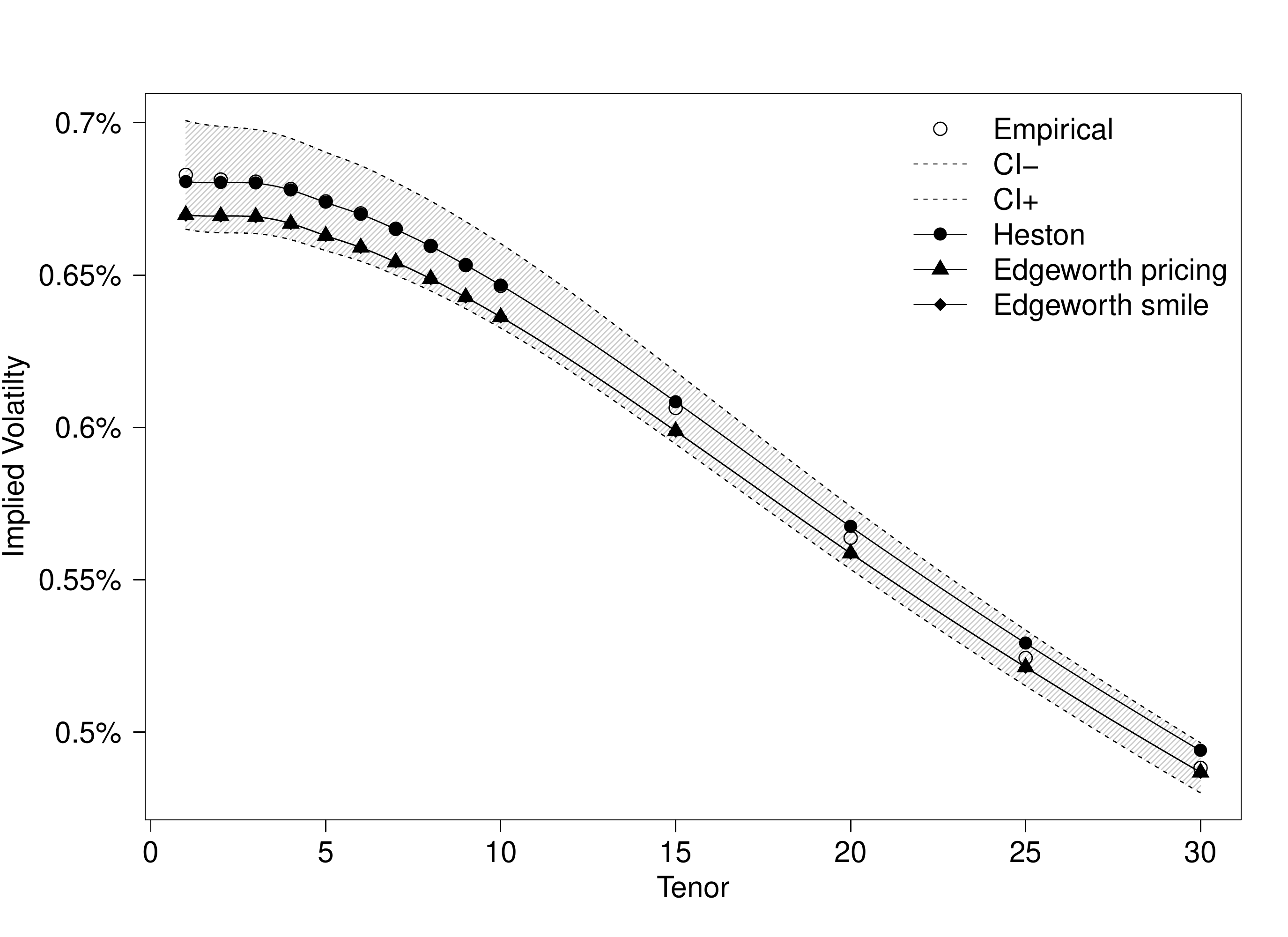}
\caption{ATM swaption volatilities with given parameters for 20-years maturity}
\label{figures/graph_20Y_5.pdf}
\end{figure}

\begin{figure}[H]
\centering
\includegraphics[scale=0.3]{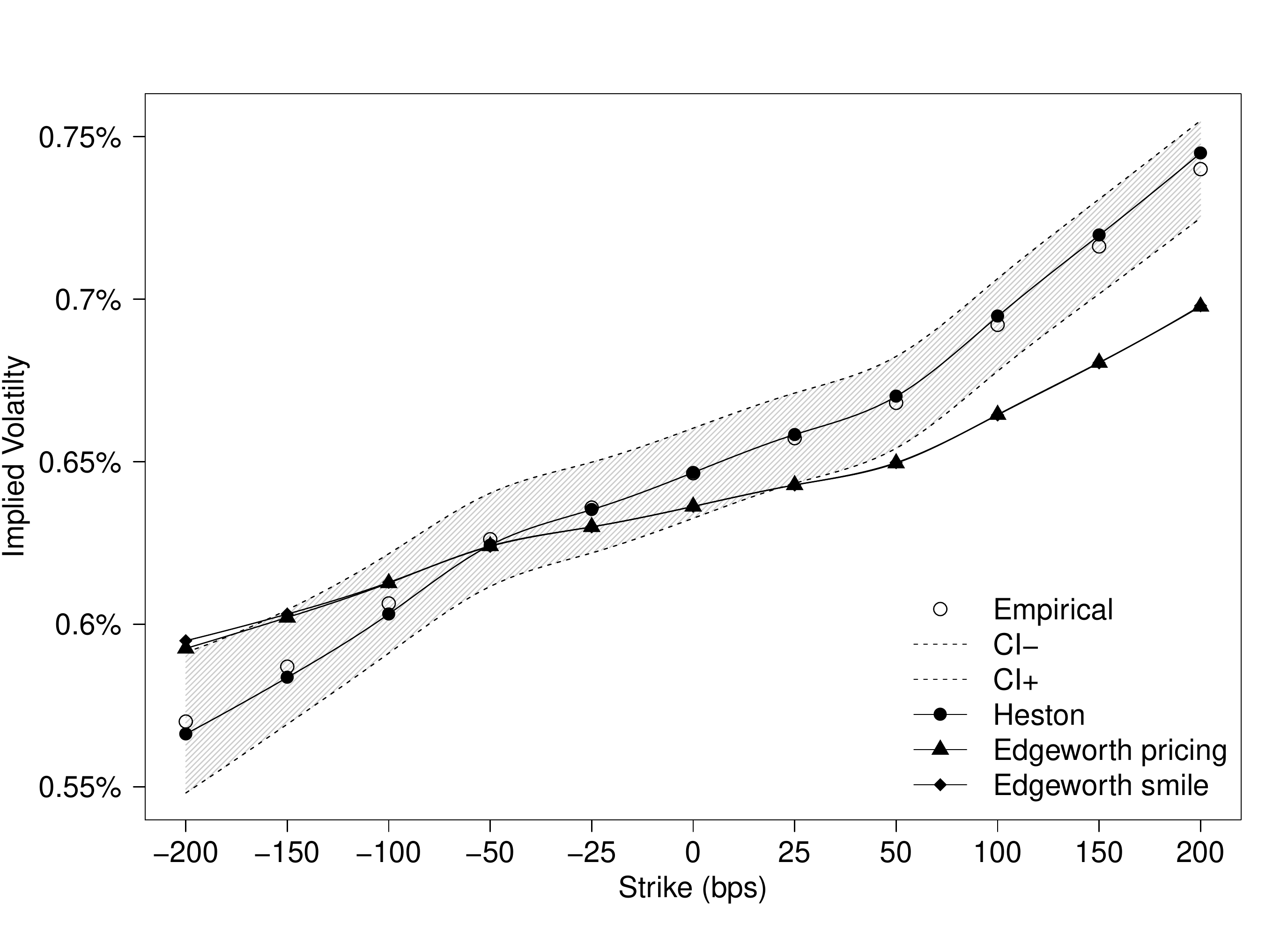}
\caption{Swaption volatility skews with given parameters for 20-years maturity}
\label{figures/graph_20Y_6.pdf}
\end{figure}

We present in this appendix the numerical results for the comparison between the Edgeworth and the Heston methods; Figures \ref{figures/graph_10Y_2.pdf} to \ref{figures/graph_10Y_6.pdf} focus on the 10-years maturity, whereas Figures \ref{figures/graph_20Y_2.pdf} to \ref{figures/graph_20Y_6.pdf} are dedicated to the 20-years maturity.
For each set, the first three figures relate to the calibration process, whereas the last two figures depict the swaption volatilities under a given set of parameters. The reader is referred to Section \ref{section_numerical} for more details.



\newpage

\bibliographystyle{ormsv080}
\bibliography{biblio}

\section*{Acknowledgements}
The authors are grateful to their colleagues at Milliman for fruitful discussions and enlightening comments, in particular Jean-Baptiste Garnier, Abdallah Laraisse, Damien Louvet and Julien Vedani.

\end{document}